\documentclass[structabstract]{aa}

\usepackage{graphicx}
\usepackage{lscape}
\usepackage{amsmath}
\usepackage{txfonts}
\usepackage{url}
\usepackage{natbib}
\usepackage[usenames,dvipsnames,svgnames,table]{xcolor}
\usepackage{ulem}

\begin{document} 

\title{Populations of rotating stars}
\subtitle{III. SYCLIST, the new Geneva Population Synthesis code}
\author{C. Georgy\inst{\ref{inst1},\ref{inst2}}, A. Granada\inst{\ref{inst3}}, S. Ekstr\"om\inst{\ref{inst3}}, G. Meynet\inst{\ref{inst3}}, R. I. Anderson\inst{\ref{inst3}}, A. Wyttenbach\inst{\ref{inst3}}, P. Eggenberger\inst{\ref{inst3}}, A. Maeder\inst{\ref{inst3}}}
\authorrunning{Georgy et al.}
\institute{Astrophysics group, EPSAM, Keele University, Lennard-Jones Labs, Keele, ST5 5BG, UK\label{inst1}\\
\email{c.georgy@keele.ac.uk}
\and
Centre de Recherche Astrophysique de Lyon, Ecole Normale Sup\'erieure de Lyon, 46, all\'ee d'Italie, F-69384 Lyon cedex 07, France\label{inst2}
\and 
Geneva Observatory, University of Geneva, Maillettes 51, CH-1290 Sauverny, Switzerland\label{inst3}\\}

\date{Received ; accepted } 
\abstract
   {Constraints on stellar models can be obtained from observations of stellar populations, provided the population results from a well defined star formation history.}
   {We present a new tool for building synthetic colour-magnitude diagrams of coeval stellar populations. We study, from a theoretical point of view, the impact of axial rotation of stars on various observed properties of single-aged stellar populations: magnitude at the turnoff, photometric properties of evolved stars, surface velocities, surface abundances, and the impact of rotation on the age determination of clusters by an isochrone fitting. One application to the cluster NGC 663 is performed.}
   {Stellar models for different initial masses, metallicities, and  zero-age main sequence (ZAMS) rotational velocities are used for building interpolated stellar tracks, isochrones, and synthetic clusters for various ages and metallicities. The synthetic populations account for the effects of the initial distribution of the rotational velocities on the ZAMS, the impact of the inclination angle and the effects of gravity and limb darkening, unresolved binaries and1 photometric errors. Interpolated tracks, isochrones, and synthetic clusters can be computed through a public web interface.}
   {For clusters with a metallicity in the range $[0.002,0.014]$ and an age between $30\,\text{Myr}$ and $1\,\text{Gyr}$, the fraction of fast rotators on the main sequence (MS) band is the largest just below the turnoff. This remains true for two different published distributions of the rotational velocities on the ZAMS. This is a natural consequence of the increase in the MS lifetime due to rotation. The fraction of fast rotators one magnitude below the turnoff also increases with the age of the cluster between $30\,\text{Myr}$ and $1\,\text{Gyr}$. The most nitrogen-rich stars are found just below the turnoff. There is an increase in the fraction of enriched stars when the metallicity decreases. We show that the use of isochrones computed from rotating stellar models with an initial rotation that is representative of the average initial rotation of the stars in clusters provides a reasonable estimate of the age, even though stars in a real cluster did not start their evolution with an identical initial rotation.} 
   {}


\maketitle

\section{Introduction}

Populations of coeval stars such as those found in many open clusters in the Galaxy and the Magellanic Clouds represent an excellent benchmark to sample of stars with identical initial composition and age. The morphology of the distribution of stars in the colour-magnitude diagrams provides both very interesting constraints on stellar models and a way to obtain the age of the population, although of course ages are still model dependent.

Our aim in the present work is twofold. On the one hand, we present the new tool SYCLIST (for SYnthetic CLusters, Isochrones, and Stellar Tracks), which was developed to produce single-aged stellar populations built on recently published grids of stellar models \citep{Ekstrom2012a,Georgy2013a,Georgy2013b}. This tool will be improved in the future, allowing non-coeval populations to be described. It is presently partially available through a web interface\footnote{\footnotesize{\url{http://obswww.unige.ch/Recherche/evoldb/index/}.}}. On the other hand, we present applications for investigating various effects. Among them we study the impact of the dispersion in the initial rotational velocities on the age determinations through the isochrone method. We also quantify the effects of gravity and limb darkening on the photometric appearance of a cluster. To our knowledge, these two darkening effects (or brightening, depending on  the inclination) are for the first time accounted for in the building of colour-magnitude diagrams (CMDs). We discuss where the fast rotators and nitrogen-rich stars are located in clusters of various ages and metallicities. A complete discussion that includes the whole parameter space (age, metallicity, velocity, and inclination distributions, presence or not of the gravity- or limb-darkening, of unresolved binaries, calibrations between effective temperature-colours and bolometric corrections, accounting for photometric noise) is beyond the scope of this paper that aims only at approaching the impact  these effects and their interplay might have on synthetic stellar populations. We eventually compare the output of the present tool with the cluster NGC 663 and show that a reasonable description of observed data can be achieved.

The SYCLIST code operates on the following stellar models libraries:
\begin{itemize}
\item the large grids of models covering most of the stellar mass domain (between $0.8$ and $120\,M_\sun$), with two different initial equatorial velocities ($V_\text{eq, ini}/V_\text{crit} = 0$ and $0.4$), and two metallicities: $Z=0.014$ \citep[solar metallicity,][]{Ekstrom2012a} and $Z=0.002$ \citep[SMC metallicity,][]{Georgy2013b};
\item the grids centred on the B-type star mass domain (between $1.7$ and $15\,M_\sun$) with nine different initial rotation rates between $\Omega_\text{ini}/\Omega_\text{crit} = 0$ and $0.95$ (where $\Omega_\text{crit} = \sqrt{\frac{GM}{R^3_\text{e, crit}}}$ and $R_\text{e, crit}$ the equatorial radius when the star rotates at the angular velocity $\Omega_\text{crit}$) at three metallicities: $Z=0.014$, $Z=0.006$ (LMC metallicity) and $Z=0.002$ \citep{Georgy2013a};
\item in addition to the two previous sets of models, the online version also includes the grid of non-rotating stellar models from \citet{Mowlavi2012a}, with a very fine mesh covering the mass domain between $0.5$ and $3.5\,M_\sun$ and $6$ metallicities between $Z=0.006$ and $Z=0.04$ .
\end{itemize}

Four different outputs are proposed\footnote{\footnotesize{Note that the online version of SYCLIST offers as of today only the first three outputs: Interpolated tracks and isochrones are directly computed online, and a request form for synthetic cluster computation can be sent through the same interface.}}:
\begin{enumerate}
\item\textit{Interpolated single stellar models:} interpolated tracks are provided for any choice of the initial mass, metallicity, and equatorial  rotational velocity. The range of allowed values are determined by the choice of the library;
\item\textit{Isochrones:} isochrones are computed with a given initial metallicity and rotational velocity;
\item\textit{Synthetic coeval stellar populations:} synthetic clusters of single-aged stellar populations are built, offering various optional settings for the initial distributions of the velocities \citep[][Dirac]{Huang2006a,Huang2010a} and of the inclination angles (random or Dirac), for the account for the gravity and limb darkening \citep{vonZeipel1924a,EspinosaLara2011a,Claret2000a}. The effect of unresolved binaries and photometric noise can also be added. Various calibrations for the transformation of the theoretical quantities (luminosities and effective temperatures) to the observed magnitudes and colours can be used (see Sect.~\ref{Sect_CalibrationCouleur});
\item\textit{Time evolution of star count:} In this mode, the code computes the evolution of the relative numbers of various types of stars (spectral types, blue-, yellow-, and red-supergiants, Wolf-Rayet subtypes, stars in a given rotation rate range, \textit{etc}.) as a function of time.
\end{enumerate}

The paper is structured as follows. In Section~\ref{Sec_Code}, we describe precisely how a synthetic population of stars is built in SYCLIST. In Section~\ref{Sec_Results}, we use the code to explore specific questions regarding mainly the impact of rotation on isochrones and synthetic population. A comparison with the observed cluster NGC 663 is made in Section~\ref{Sec_NGC663}. Finally, our conclusions are presented in Section~\ref{Sec_Conclu}.

\section{Description of the population synthesis code}\label{Sec_Code}

Whatever the chosen output, the SYCLIST code relies on its ability to interpolate between existing stellar tracks. For this purpose, the electronic tables provided by the Geneva group (and used as input libraries in SYCLIST) are formatted as follows \citep[see][for more details]{Ekstrom2012a}: 400 points (hereafter time-models) are extracted from a complete computed sequence of a few tens of thousands time-models. These 400 time-models are chosen carefully in order to fully describe the morphology of the tracks in the theoretical Hertzsprung-Russell diagram (HRD) ensuring that reasonable interpolations may be made between time-models having the same number in these tables.

\subsection{Interpolated stellar tracks}

In order to obtain an evolutionary track for a star with a given initial metallicity, mass and rotational rate $\omega$, we perform linear interpolations\footnote{\footnotesize{Note that for quantities such as the mass, time, effective temperature and luminosity, the interpolation is linear in the \textit{log space}.}}, first in $\omega$, then in mass and finally in metallicity, using the 8 closest surrounding models in the chosen database. The models by \citet{Ekstrom2012a}, and \citet{Mowlavi2012a} can be interpolated among various masses and metallicities. In addition, interpolation in $\omega$ is available for the grids by \citet{Georgy2013a}, since these feature a dense coverage of initial rotation rates.

In case an \textit{extrapolation}\footnote{\footnotesize{SYCLIST performs \textit{extrapolations}, if models with initial rotation rates higher than $0.95$ (maximal $\omega$ in \citet{Georgy2013a} are requested.}} has to be performed, and for quantities that have a limited range of possible values or a threshold, such as the chemical abundances, the rotation rate, the mass-loss rates, \textit{etc.}, the interpolated values are constrained to not overstep the limit (for example, if the interpolated value of the central hydrogen abundance is $-0.003$, this value is reset to $0.0$).

If the SYCLIST code is launched in the mode ``interpolated stellar model", an output is generated at this step and is formatted to be identical to the electronic tables described in \citet{Ekstrom2012a}. This tool is useful to attribute a mass and an age to a single star whose position in the Hertzsprung-Russell diagram (HRD) or in the $M_\text{V}$ versus B-V plane is known.

Obviously, an interpolated model represents only an approximation of a rigorously computed stellar model. In Section~\ref{Sec_StellarModels}, we present examples of interpolated tracks and discuss their reliability when confronted to the computed tracks.

\subsection{Isochrone building}

An isochrone at a given age $t$ is computed by interpolating a set of stellar models with initial mass in the range $[M_\text{min},M_\text{max}]$ with $M_\text{min}$ the minimal mass in the models library and $M_\text{max}$ the maximal mass of the stars that are still in a nuclear burning phase at time $t$. In a second step, all the pertinent quantities are obtained by performing for each interpolated model a new linear interpolation along the $\log(t)$ axis.

In ``Isochrone'' mode, the outputs of SYCLIST provide all the surface quantities of the stellar models ($\log(L/L_\sun)$, $\log(T_\text{eff})$, magnitude, colours, abundances, rotation characteristics,\dots).

\subsection{Synthetic clusters\label{SectSynthClust}} 

\begin{figure}
\centering
\includegraphics[width=0.47\textwidth]{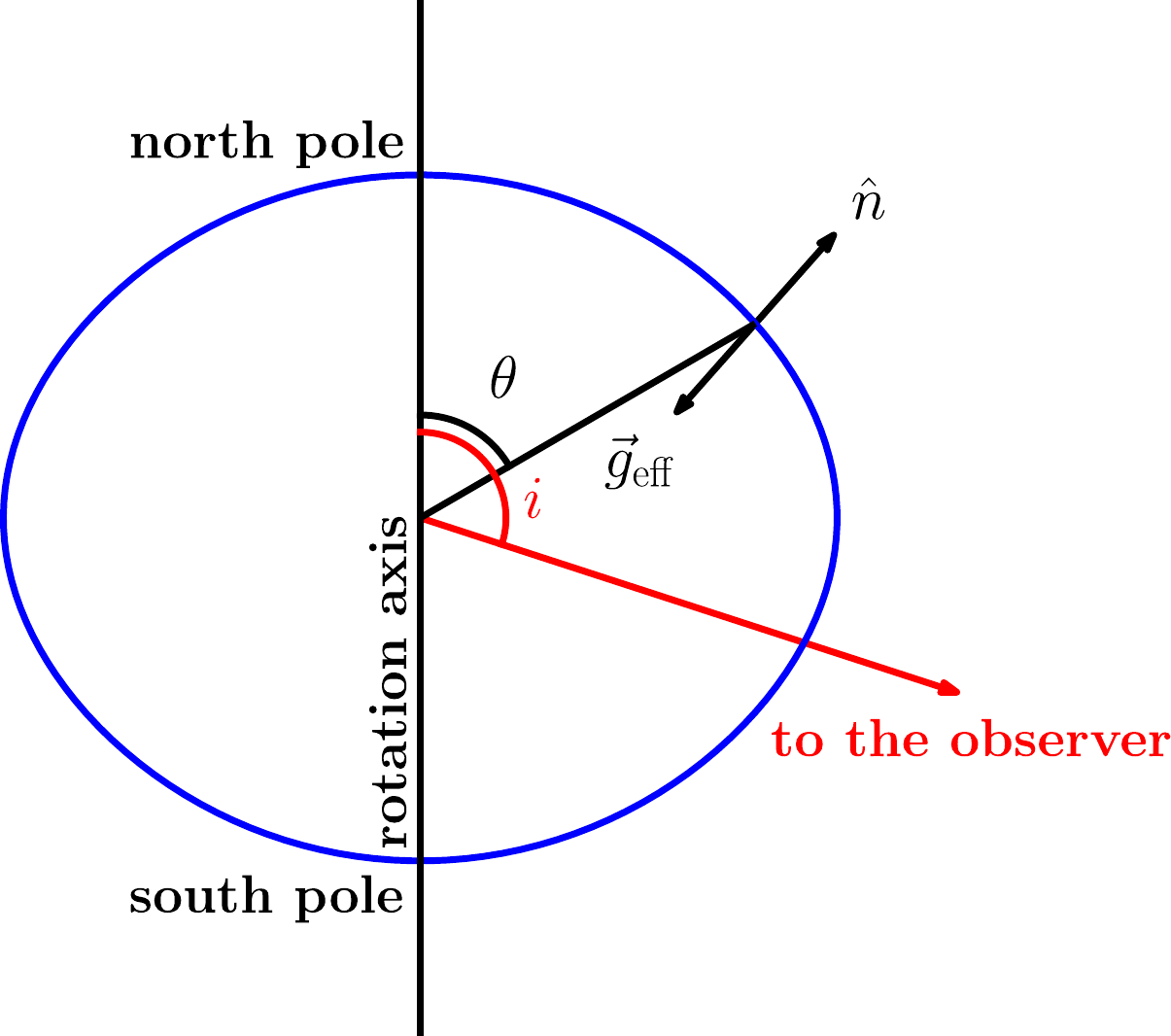}
\caption{Schematic representation of a meridional cut of a rotating star. The plane shown contains the line connecting the centre of the star to the observer and the rotational axis. In the text, the angle $i$ is called the inclination angle ($i=0$ when the star is seen pole-on). A given point on the surface is defined by the angle $\theta$ that we call the colatitude in the text. Note that in a deformed rotating star, the unit vector perpendicular to the surface $\hat{n}$ is not aligned with the direction connecting the considered position to the centre of the star. The same is true for the effective gravity which is the classical gravity modified by the effect of the centrifugal force.}
\label{schema}
\end{figure}

To build synthetic clusters, we assume that each star is characterised by the following physical properties: its initial mass, metallicity, and rotational velocity, the inclination axis between the axis of rotation and the direction of the observer ($i=0$ in case of a star seen pole-on, see Fig.~\ref{schema}), its age, the presence or absence of an unresolved companion\footnote{\footnotesize{This list is obviously not exhaustive, and we consider here only the physical properties that we implemented in our grids of stellar models or in the SYCLIST code. Other physical processes, such as internal magnetic fields, or surface magnetic braking, could be added by using specific stellar tracks libraries.}}. For each star of the cluster, the initial mass, rotational velocity, and inclination angle is determined through a random draw performed in such a way that it follows the desired distribution\footnote{\footnotesize{Technically, the draw is made in two steps. The first is a uniform draw between $0$ and $1$, the second step is then to transform it into the desired quantity by inverting the corresponding cumulative distribution function.}}. In addition, calibration functions are needed to translate theoretical quantities (luminosity and effective temperature) into observed ones (magnitude and colour), and some uncertainties in the magnitudes and colours might be accounted for. In the following, we explain all the options that are implemented in the SYCLIST code for these settings.

\subsubsection{The initial mass function, the initial distribution of rotational velocities and of inclinations}

In the current version of SYCLIST, only the initial mass function (IMF) from \citet{Salpeter1955a}, described by a power-law function with an index $\alpha = -2.35$, has been implemented. Even though Salpeter's work is almost 60 years old, newer and more complex IMFs found by other authors still show that this index is valid in the range of stellar masses considered in this work \citep[see ][]{Kroupa2002a}. This justifies the use in the present work of the simplest expression by \citet{Salpeter1955a}. In the future, the code will offer a choice of various IMFs.

Four distributions of rotational velocities are currently implemented:
\begin{enumerate}
\item the simplest distribution, is a Dirac distribution $\delta(\omega)$, with $\omega$ between $0$ and $1$. The quantity $\omega$ is equal to $\Omega_\text{ini}/\Omega_\text{crit}$ where $\Omega$ is the surface angular velocity. With this distribution, it is possible to build a synthetic cluster (or isochrones) in which all the stars have the same $\omega$ on the ZAMS. This simple distribution is very practical to study the effect of rotation in stellar clusters, in particular when rapid rotators are considered;
\item a uniform distribution with $\omega$ between 0 and 1 is also implemented. Although this distribution is certainly not a physical one, it allows to study the effect of the other parameters change, like the change of the angle of view of the star and also the effect of the age on the angular velocity distribution;
\item from the literature, we get the \citet{Huang2006a} distribution, established from the observation of 496 galactic OB stars of the field and clusters of different ages;
\item \citet{Huang2010a} propose a distribution of rotational velocities for B stars. By measuring the projected velocity $V\sin(i)$ of 220 young galactic B stars, they made a simple polynomial fit of the histogram data of $V\sin(i)/V_\text{crit}$  and then used a deconvolution algorithm to derive the distribution of $V_\text{eq}/V_\text{crit}$. They obtained a rotational velocity distribution for young stars for three ranges of masses: low mass ($2\,M_\sun \leq M < 4\,M_\sun$), middle mass ($4\,M_\sun\leq M < 8\,M_\sun$) and high mass range ($M\geq 8\,M_\sun$).
\end{enumerate}

In the last two cases, we converted $V_\text {eq}/V_\text{crit}$ to $\omega$, because this quantity is the one used to build the grid of stellar models with nine different initial rotation rates \citep{Georgy2013a}. In the simple Roche model assumed to compute the shape of our stellar surfaces, there is a unique relation between the ratio $V_\text{eq}/V_\text{crit}$ and $\omega$ \citep[see for instance][]{Maeder2009a}, that we use here to convert one quantity into the other one.

For rotating stars, in particular if they are fast rotators, the inclination angle under which the star is seen will affect the observed temperature and luminosity due to gravity darkening. The angle of view is  drawn from the following distributions:
\begin{enumerate}
\item a Dirac one, with an open choice for the angle;
\item  a $\sin(i)$ probability;
\end{enumerate}

\subsubsection{Gravity-darkening effect and impact of inclination\label{Sec_GravDark}}

The luminosity that we obtain from our evolutionary tracks, $L_\text{MOD}$, is the bolometric power emitted by the star in all directions. We define the effective temperature given by our models by $T_\text{eff,MOD}= (L_\text{MOD}/(\sigma \Sigma))^{1/4}$, where $\sigma$ is the Stefan-Boltzmann constant and $\Sigma$ the actual total surface of the star. 

$L_\text{MOD}$ is a theoretical quantity. Observations give access to the power emitted by the star in only one direction, that of the observer. In case the star emits radiation isotropically, a measurement along any direction is equivalent, and, from the measurement in one direction, the total luminosity of the star can be correctly estimated. However, when the emission is not isotropic, not all directions are equivalent. In this case, from a measurement in a given direction, the estimation of  the luminosity, assuming isotropy, $L_\text{MES}$, is different from $L_\text{MOD}$. 

$T_\text{eff,MOD}$ as defined above is the theoretical surface averaged effective temperature. In case the star presents variations of its temperature over its surface, the integration of the local radiative flux (proportional to $\sigma T_\text{eff}^4$ where $T_\text{eff}^4$ is the local effective temperature) would still be equal to $L_\text{MOD}$. Observationally, the effective temperature of a star is obtained by spectroscopy or through colours that are measured on the hemisphere facing the observer. Again, in case of isotropy, any hemisphere is equivalent and such a measurement provides the effective temperature of the star. In case of anisotropy, one has only access to $T_\text{eff,MES}=(L_\text{MES}/(\sigma \Sigma_\text{p}))^{1/4}$, where $\Sigma_\text{p}$ is the projected surface of the star on a plane perpendicular to the line of sight. In order to generate observational quantities, we first need to link $L_\text{MES}$ to $L_\text{MOD}$ and $T_\text{eff,MES}$ to $T_\text{eff,MOD}$.

There is evidence indicating that the surface of a rotating star is, at least as a first approximation, well described by the Roche Model \citep[e.g.][]{DomicianoDeSouza2003a}. In this model, the ratio of the equatorial radius to the polar radius of the star depends on $\omega$, and is greater for more rapidly rotating stars. For the most rapidly rotating stars ($\omega > 0.7$), the equatorial radius becomes significantly larger than the polar radius. As a consequence, the effective gravity of rotating stars with a radiative surface is latitude dependent. Because there is a relation between $g_\text{eff}$ and $T_\text{eff}$ \citep{vonZeipel1924a,EspinosaLara2011a}, the effective temperature is also latitude dependent \citep[see \textit{e.g.}][]{Maeder1999a}. The poles of a rotating star are hotter than the equatorial regions, which turn out to be cooler and less bright. This temperature and brightness contrast between the poles and the equator of a rotating star is known as gravity darkening.

Due to gravity darkening, a star seen pole-on will be observed as hotter and more luminous (bluer and brighter) than it would be if seen equator-on. Therefore, the angle of view under which we observe a star will influence its location in the HRD, by modifying the inferred luminosity and effective temperature.

The luminosity inferred by an observer in a direction $\mathbf{d}$ inclined by an angle $i$ with respect to the rotation axis is:
\begin{equation}
L_\text{MES}(i) = 4\pi\int_{\text{d}\mathbf{\Sigma}\cdot\mathbf{d} > 0} I(\theta)\,\text{d}\mathbf{\Sigma}\cdot\mathbf{d},\label{Eq_Lumi}
\end{equation}
with $I(\theta)$ the specific intensity at the colatitude $\theta$. The integral is computed on the hemisphere that is visible for the observer, \textit{i.e.} the points of the stellar surface with $\text{d}\mathbf{\Sigma}\cdot\mathbf{d} > 0$. Assuming a black body radiation, this leads to:
\begin{equation}
L_\text{MES}(i) = 4\sigma\int_{\text{d}\mathbf{\Sigma}\cdot\mathbf{d} > 0} T_\text{eff}^4(\theta)\,\text{d}\mathbf{\Sigma}\cdot\mathbf{d}.
\label{LMes}
\end{equation}

$L_\text{MES}$ is thus sensitive to the way the effective temperature is distributed over the surface. In the SYCLIST code, two relations between $T_\text{eff}(\theta)$ and the colatitude are implemented:
\begin{itemize}
\item  a formulation based on the von Zeipel theorem \citep{vonZeipel1924a}: in the framework of the Roche model, we have that
\begin{equation}
T_\text{eff}(\theta)=T_\text{eff,MOD}\,\left( \frac{g_\text{eff}(\theta)}{\bar{g}_\text{eff}} \right)^{1/4};
\end{equation}
\item the formulation proposed by \citet{EspinosaLara2011a}: their Eq. (31) can be re-written as
\begin{align}
T_\text{eff}(\theta)=&T_\text{eff,MOD}\,\left( \frac{\Sigma}{4\,\pi\,R_\text{p,crit}^2} \right)^{1/4}\,\left[ \left( -\frac{1}{x^2}+\frac{8}{27}\omega^2x\,\sin^2(\theta) \right)^2\right. \notag\\
&\left. + \left( \frac{8}{27}\omega^2x\,\sin(\theta)\cos(\theta) \right)^2 \right]^{1/8}\,\sqrt{\frac{\tan(\vartheta)}{\tan{\theta}}}
\end{align}
where $x=R(\theta)/R_\text{p,crit}$ (with $R_\text{p,crit}$ the polar radius at the critical velocity), and $\vartheta$ the solution of their Eq.~(24).
\end{itemize}
In both cases, $T_\text{eff}(\theta)$ can be expressed as:
\begin{equation}
T_\text{eff}(\theta)=T_\text{eff,MOD}\,f(x,\omega,\theta),
\label{eq_Tefftheta}
\end{equation}
In the Roche model approximation, the shape of the stellar surface (given by the function $x$) only depends on $\omega$ \citep[\textit{e.g.}][]{Maeder2009a}. Thus, the function $f(x,\omega,\theta)$ is independent from the stellar parameters and only depends on geometrical considerations.

Inserting Eq.~\ref{eq_Tefftheta} in Eq.~\ref{LMes} yields:
\begin{equation}
L_\text{MES}(i)= L_\text{MOD}\,\frac{4}{\Sigma}\,\int_{\text{d}\mathbf{\Sigma}\cdot\mathbf{d} > 0} f^4(x,\omega,\theta)\,\text{d}\mathbf{\Sigma}\cdot\mathbf{d}.
\end{equation}
Thus, the observed luminosity of the star depends on the real luminosity corrected by a purely geometrical factor $C_L(i,\omega)$:
\begin{equation}
C_L(i,\omega) = \frac{4}{\Sigma}\int_{\text{d}\mathbf{\Sigma}\cdot\mathbf{d} > 0} f^4(x,\omega,\theta)\,\text{d}\mathbf{\Sigma}\cdot\mathbf{d}.\label{Equ_CL}
\end{equation}
This factor is independent of the stellar parameters\footnote{\footnotesize{The stellar parameter-dependent surface vector $\text{d}\mathbf{\Sigma}$ is canceled by the prefactor $\Sigma^{-1}$.}}.

The observed $T_\text{eff, MES}$ is deduced by averaging the flux on the \textit{projected} stellar surface $\Sigma_\text{p}$:
\begin{equation}
T_\text{eff, MES}^4(i,\omega) = \frac{1}{\Sigma_\text{p}}\int_{\text{d}\mathbf{\Sigma}\cdot\mathbf{d} > 0}T_\text{eff}^4(\theta)\,\text{d}\mathbf{\Sigma}\cdot\mathbf{d}.
\end{equation}
Again with the help of Eq.~(\ref{eq_Tefftheta}), we have:
\begin{equation}
T_\text{eff, MES}(i,\omega) = T_\text{eff, MOD}\left[\frac{1}{\Sigma_\text{p}}\int_{\text{d}\mathbf{\Sigma}\cdot\mathbf{d} > 0} f^4(x,\omega,\theta)\,\text{d}\mathbf{\Sigma}\cdot\mathbf{d}\right]^\frac{1}{4}.
\end{equation}
Hence, also the observed effective temperature is a function of average effective temperature, corrected by the factor $C_{T\text{eff}}$:
\begin{equation}
C_{T\text{eff}}(i,\omega) = \left[\frac{1}{\Sigma_\text{p}}\int_{\text{d}\mathbf{\Sigma}\cdot\mathbf{d} > 0} f^4(x,\omega,\theta)\,\text{d}\mathbf{\Sigma}\cdot\mathbf{d}\right]^\frac{1}{4}.\label{Equ_CT}
\end{equation}
Both $C_{\text{L}}$ and $C_{T\text{eff}}$ are purely geometrical, and only depend on $\omega$ and $i$. One can therefore compute them once for various inclinations and rotations, and use these values for performing interpolations.

\subsubsection{Limb-darkening effect and impact of inclination}

The gravity-darkening effect discussed above arises because each of the surface element has its own effective temperature. Considered alone, it is supposed that the specific intensity emitted by a surface element in a given direction does not depend on that direction. However, the atmosphere of a real star does not radiate as a true black body: the specific intensity will not be the same when the elementary surface is seen face-on or from a grazing viewing angle. Due to the structure of the atmosphere, the optical thickness is not the same along these two directions and thus the radiation received will come from more or less deep (and thus more or less warm) layers of the star. This is called the limb-darkening effect. For a uniformly bright star, it results in a  dimming of the star's disk from the centre towards the edge. To reproduce the emission from the hemisphere oriented towards us, we have to integrate the specific intensity arising from directions perpendicular to the surface in the centre of the hemisphere and directions tangent to the surface at the edge, with all the intermediate values in-between.

Let us first describe the equations used for a non-rotating star whose surface is characterised by a single value of the effective temperature. The limb-darkening law is expressed by $\ell(\mu)=I(\mu)/I(1)$, where $I$ is the specific intensity and $\mu=\frac{\text{d}\Sigma\cdot\mathbf{d}}{|\text{d}\Sigma\cdot\mathbf{d}|}$, the cosine of the angle between the normal to the surface and the direction towards the observer. A value of $\mu=1$ corresponds to the centre of the stellar disk, whereas $\mu=0$ to the stellar edge or limb. 

To obtain the impact of this effect on the perceived luminosity we can use Eq.~(\ref{Eq_Lumi}). Since the case considered here is the one of a non-rotating star, we have that the perceived luminosity without any limb-darkening effect would be
\begin{equation}
L_\text{MES}= 4\pi\int_{\text{d}\mathbf{\Sigma}\cdot\mathbf{d} > 0} I \text{d}\mathbf{\Sigma}\cdot\mathbf{d}.
\end{equation}
The limb-darkening effect implies that the specific intensity depends on $\mu$. Since $I(\mu)=\ell(\mu) I(1)$, one can write
\begin{equation}
L= 4\pi\int_{\text{d}\mathbf{\Sigma}\cdot\mathbf{d} > 0} \ell(\mu) I(1) \text{d}\mathbf{\Sigma}\cdot\mathbf{d}.\label{eq_lum}
\end{equation}
To proceed, one needs to find a relation between $I(1)$ and the local effective temperature of the surface element considered. For this purpose we can use  the fact that the mean specific intensity $\left<I\right>$  is related to the effective temperature via the relation
\begin{equation}
\left<I\right>= 2\int_{0}^{\pi/2}  I \cos(\alpha) \sin(\alpha) \text{d}\alpha=\frac{\sigma T_\text{eff}^4}{\pi}.
\end{equation}
It can be shown using the definition of $\ell(\mu)$ that
\begin{equation}
\frac{I(1)}{\left<I\right>}={1 \over 2 \int_0^1 \ell(\mu) \mu \text{d} \mu}.
\end{equation}
Thus, replacing $I(1)$ by $\frac{I(1)}{\left<I\right>} \left<I\right>$ in Eq.~(\ref{eq_lum}), one has finally
\begin{equation}
L_\text{MES}=\frac{2\sigma}{\int_0^1 \ell(\mu) \mu \text{d} \mu}\int_{\text{d}\mathbf{\Sigma}\cdot\mathbf{d} > 0} \ell(\mu) T_\text{eff}^4  \text{d}\mathbf{\Sigma}\cdot\mathbf{d}.\label{Eq_FinalLum}
\end{equation}
In case there is no limb-darkening, then $\ell(\mu)=1$ and one obtains
\begin{equation}
L_\text{MES}=4\sigma \int_{\text{d}\mathbf{\Sigma}\cdot\mathbf{d} > 0} T_\text{eff}^4\text{d}\mathbf{\Sigma}\cdot\mathbf{d}.
\end{equation}
Since $I$ is equal to $\sigma T_\text{eff}^4/\pi$, then one finds again Eq.~(\ref{Eq_Lumi}).

Following \citet{Claret2000a}, we represent the quantity $\ell(\mu)$ by
\begin{equation}
 \ell(\mu)= 1 -\sum_{n=1}^{4}a_{n}(1-\mu^{n/2}),
\end{equation}
where the coefficients $a_{n}$ depend on the effective temperature and surface gravity of the star. In the present work, we use the bolometric limb-darkening coefficients from the new grids of \mbox{ATLAS9} models \citep{Howarth2011a}, which are given for a wide range of surface gravities and effective temperatures.

In the case of a rotating star, the surface gravity and $T_\text{eff}$ change with the colatitude $\theta$. Therefore, in Eq.~(\ref{eq_lum}) the latitude-dependent temperature should be accounted for in the computation of the local $I(1)$ and $\left<I\right>$. However, in the range of $T_\text{eff}$ and $\log(g)$ values typical for intermediate and massive stars, $\ell(\mu)$ does not vary much \citep[according to the tables provided by][]{Howarth2011a}. Eq.~(\ref{Eq_FinalLum}) thus remains valid. In this case, we take  the latitude-dependent temperature of the star into account. However, we take representative values for the limb-darkening coefficients for the whole star, given by the mean $T_\text{eff}$ and $\log(g)$. In other words, we assume that $I(1)/\left<I\right>$ is constant over the surface.

This approach introduces only small errors: for temperatures higher than $\sim12000\,\text{K}$, $\ell(\mu)$ remains almost constant with $T_\text{eff}$ and varies very little with $\log(g)$. For temperatures lower than $\sim12000\,\text{K}$, $\ell(\mu)$ is almost independent of log(g) for all $\mu$, albeit with a somehow stronger dependency on $T_\text{eff}$: in this temperature domain, using $T_\text{eff}$ and a representative $\log(g)$ for the most rapidly rotating stars introduces an error in $\ell(\mu)$ less than 10\% for $\mu=0.5$ and less than 15\% for $\mu=0.1$. We have to bear in mind that the current limb-darkening coefficients have been obtained for non rotating models. However, as far as we know, there does not exist a study with complete data for a broad range of $T_\text{eff}$ and $\log(g)$ for rotating models.

Then, for our rotating models with $L_\text{MOD}$ and $T_\text{eff,MOD}$, we obtain the limb-darkening coefficients for such a temperature and for a value of $\log(g)=4\pi GM\sigma T_\text{eff,MOD}^4/L_\text{MOD}$. By the mean of Eq.~(\ref{Eq_FinalLum}), one obtains $L_\text{MES}$. $T_\text{eff,MES}$ is obtained by $L_\text{MES} = \Sigma_\text{p}T_\text{eff,MES}^4$.

\begin{figure*}
\centering
\includegraphics[width=0.35\textwidth]{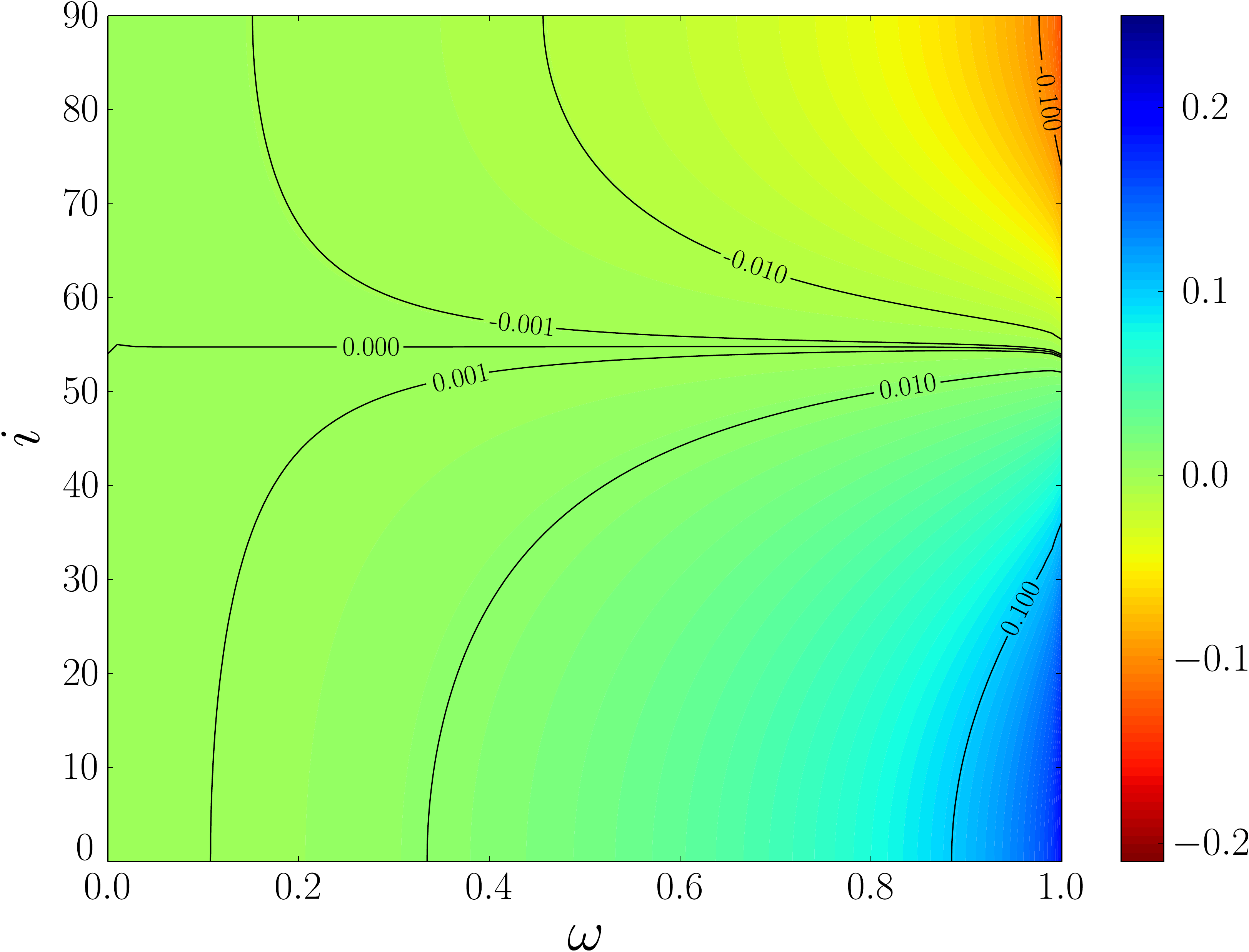}\includegraphics[width=0.35\textwidth]{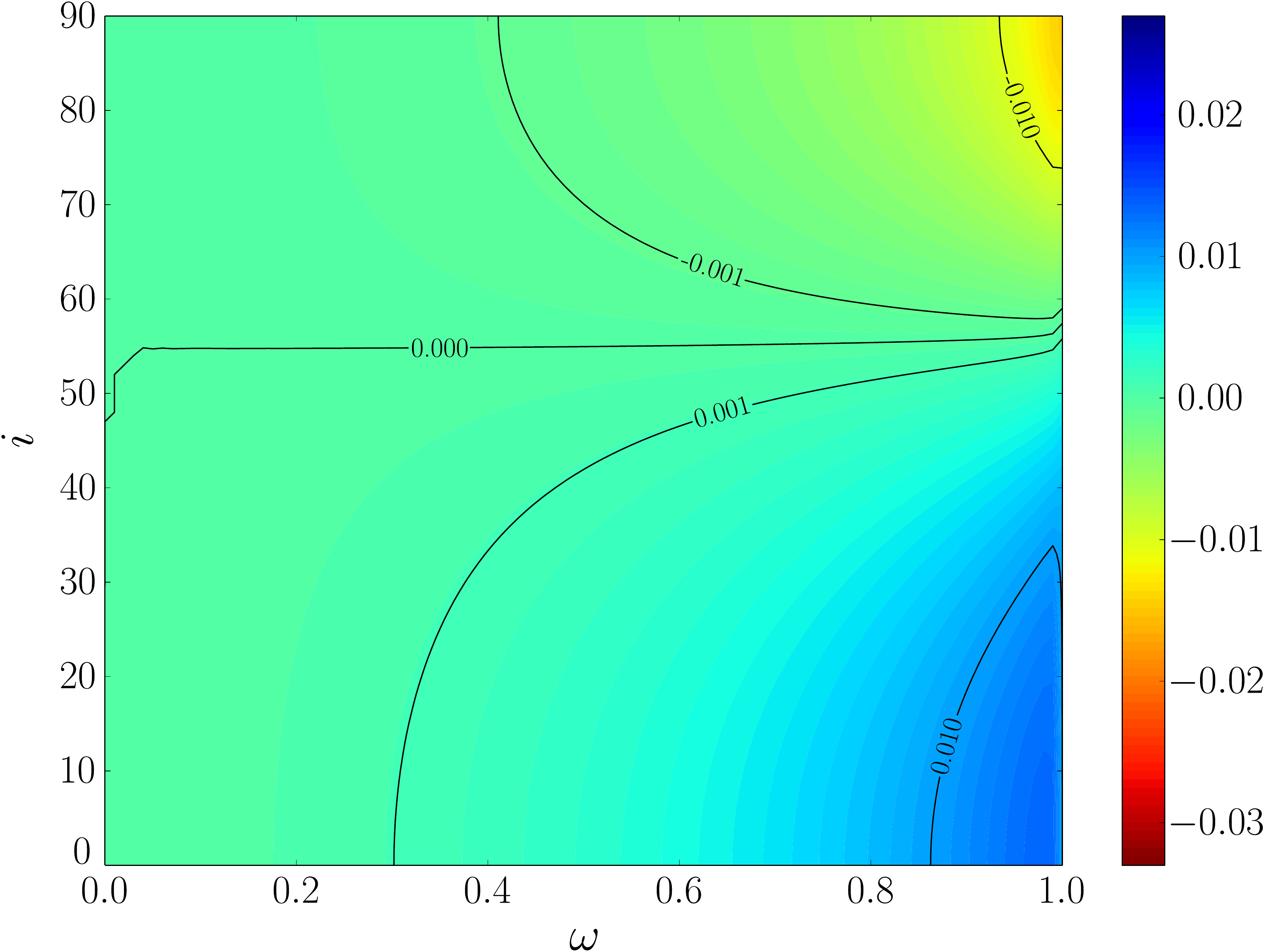}
\includegraphics[width=0.35\textwidth]{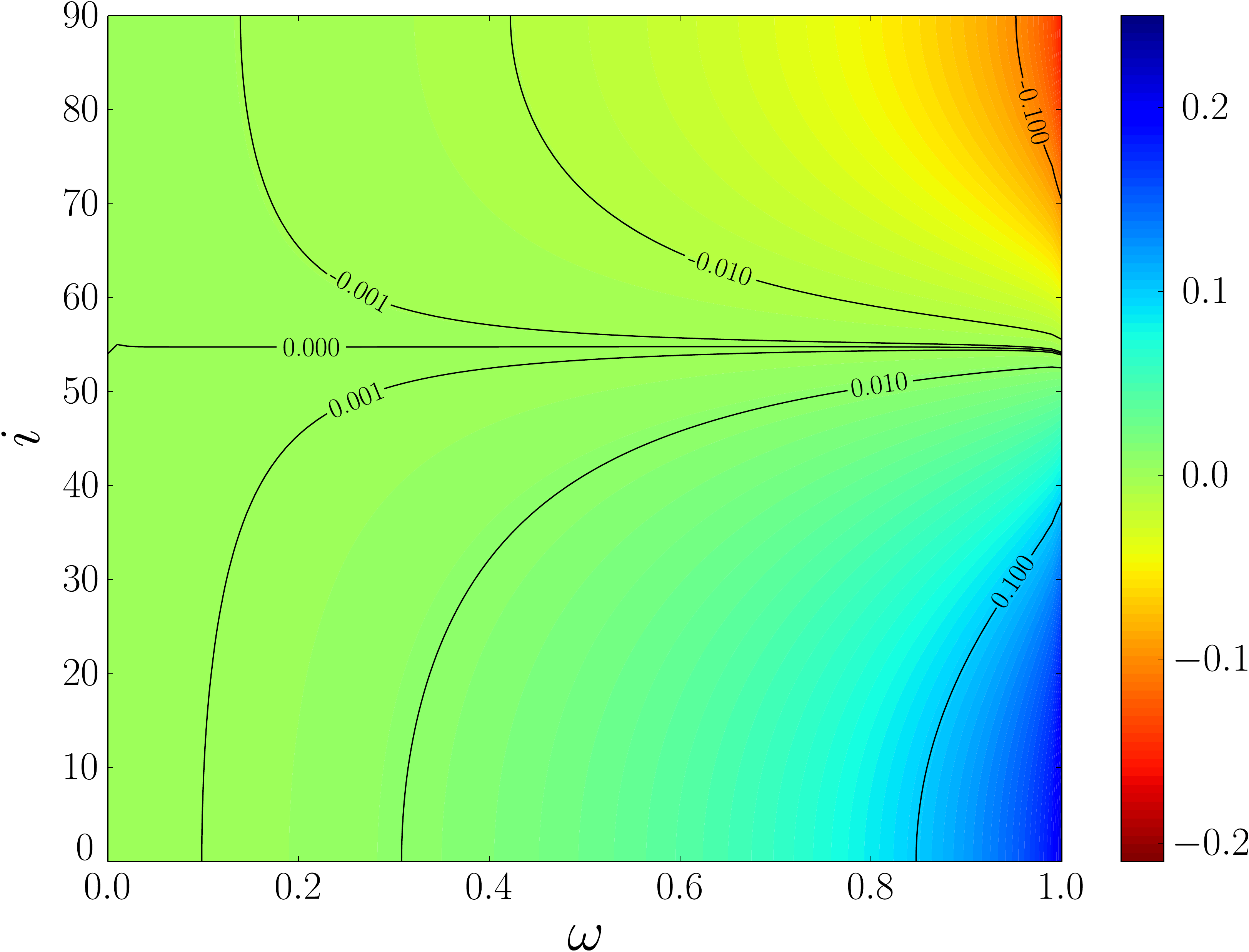}\includegraphics[width=0.35\textwidth]{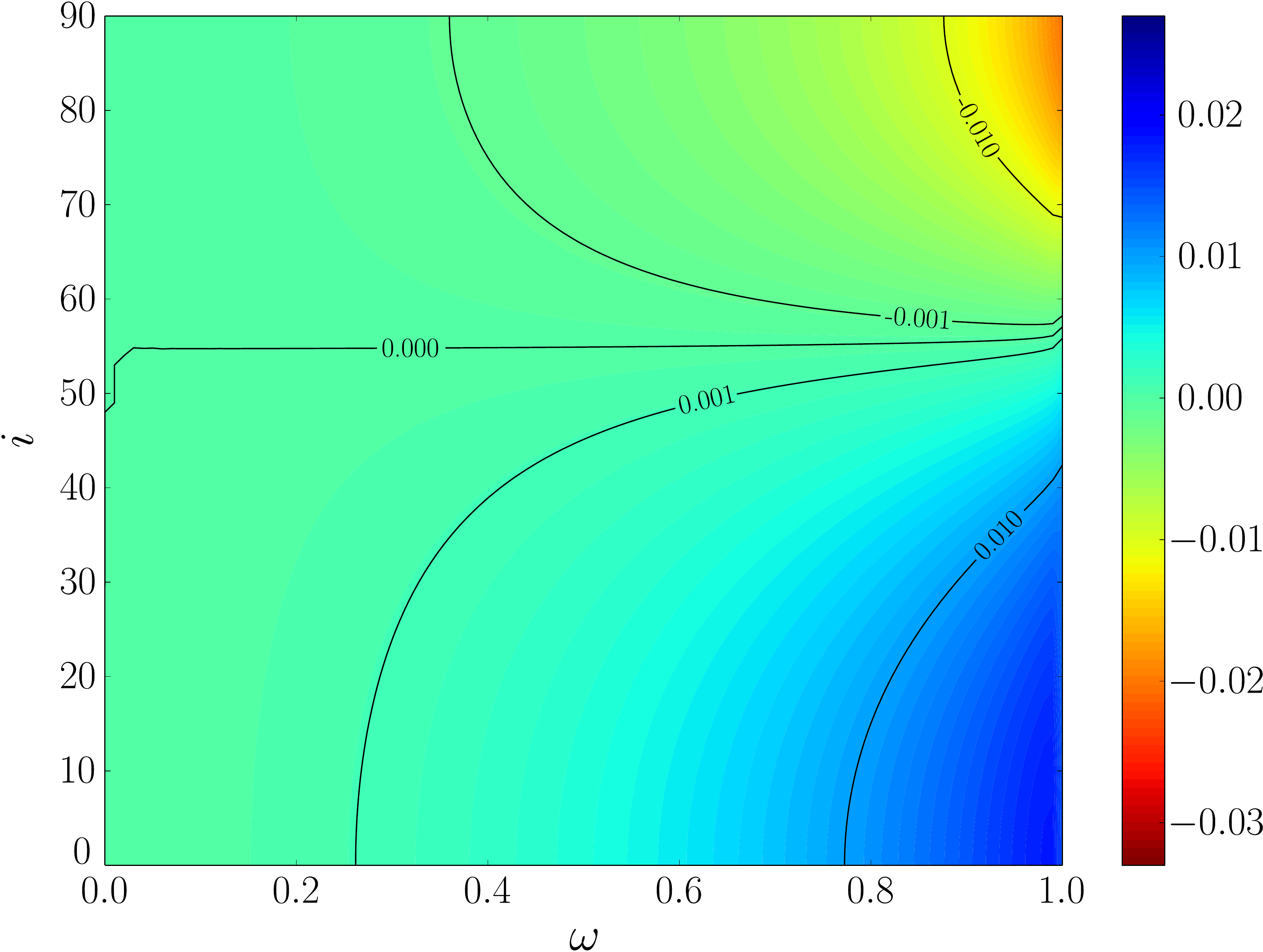}
\includegraphics[width=0.35\textwidth]{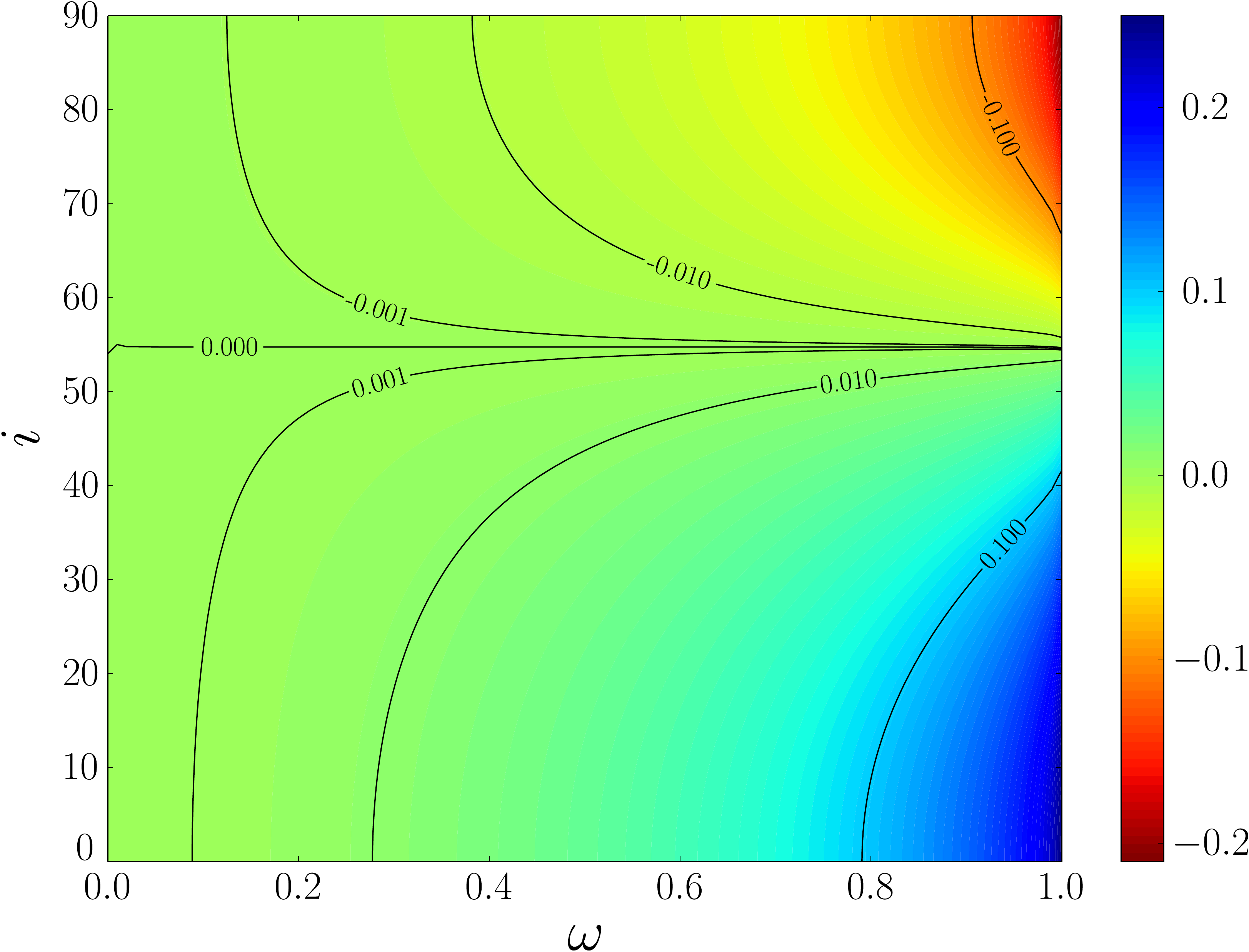}\includegraphics[width=0.35\textwidth]{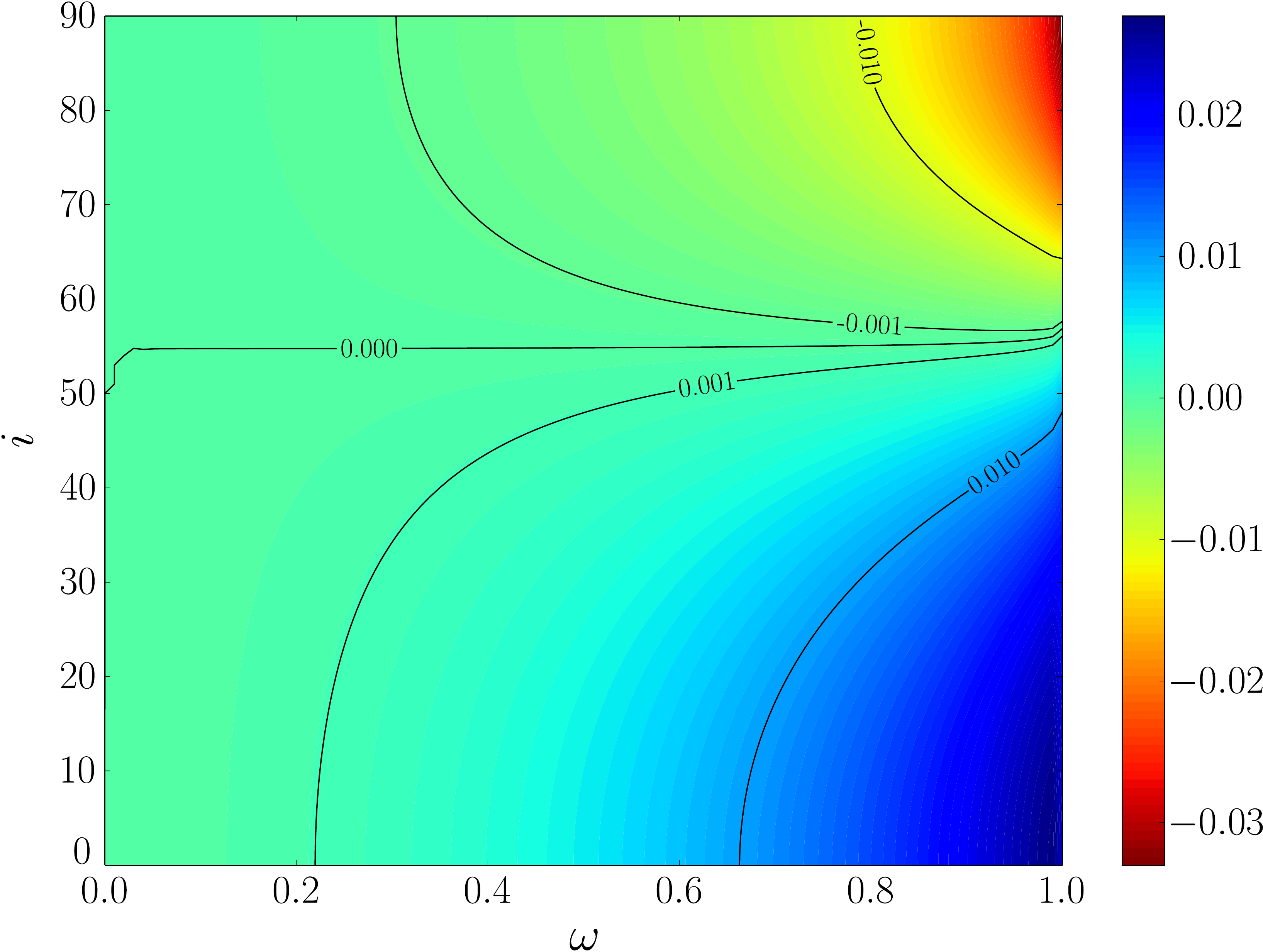}
\caption{Logarithm of the correcting factors for the effective temperature (\textit{left panels}) and luminosity (\textit{right panels}) as a function of $\omega$ and of the inclination angle $i$, with the gravity-darkening law by \citet{EspinosaLara2011a}. \textit{Top panels:} Gravity-darkening alone. \textit{Centre panels:} Gravity- and limb-darkening for $T_\text{eff} = 25 000\,\text{K}$ and $\log(g)=3.5$. \textit{Bottom panels:} same as middle panels for $T_\text{eff} = 5000\,\text{K}$ and $\log(g)=0.0$.}
\label{CORR}
\end{figure*}

Figure~\ref{CORR} shows the value of the correcting factors in dex in case only gravity darkening is accounted for and in case both gravity darkening and limb darkening are taken into account. With the temperature profile given by \citet{EspinosaLara2011a}, a star seen pole-on becomes brighter by at most $0.2$ dex, which represents an increase in about $50\%$ in luminosity. The effective temperature appears to be higher by about 0.01 dex ($2\%$). When the star is seen equator-on, the luminosity and the effective temperature are decreased: $-0.13$ dex ($\sim -25\%$) and -0.01 dex ($-3\%$) respectively. Also the limb-darkening effect reinforces the effect of gravity darkening particularly for fast rotation, and for stars seen pole- or equator-on. For instance, in the example presented in Fig.~\ref{CORR} for an object with $T_\text{eff} = 25 000\,\text{K}$ and $\log(g)\,=3.5$ seen pole-on, the correction factors for luminosity and effective temperature become $0.21$ dex ($60\%$) and $0.01$ dex ($3\%$), respectively. For the same object seen equator-on, these values become $-0.15$ dex ($-30\%$) and $-0.02$ dex ($-5\%$). Accounting for gravity-darkening alone (with either \citealt{vonZeipel1924a} or \citealt{EspinosaLara2011a} prescription), or gravity- and limb-darkenening together is optional in SYCLIST.

\subsubsection{Calibration functions}\label{Sect_CalibrationCouleur}

A colour-$T_\text{eff}$ calibration and bolometric correction is required in order to transform our theoretical HRD to a CMD. SYCLIST offers two different choices:
\begin{enumerate}
\item the calibrations used in the old grids \citep{Schaller1992a}: for the MS stars (luminosity class V), the calibration relation between the effective temperature $T_\text{eff}$ and the colour index B-V is taken from \citet{Boehm-Vitense1981a}, the bolometric correction from \citet{Malagnini1986a} and the UBV relation from \citet{Schmidt-Kaler1982a}; for giant type III and supergiant type Ia star, the  $T_\text{eff}$ versus B-V relation and the bolometric correction are taken from \citet{Flower1977a}, and the UBV relation from \citet{Schmidt-Kaler1982a};
\item the newer colour-temperature calibration for stars proposed by \citet{Worthey2011a}.
\end{enumerate}

In this work, the colours are computed as a function of the averaged $T_\text{eff,MES}$.

\subsubsection{Binary population and photometric noise}

The available libraries rely on single star tracks at the moment and thus do not account for the evolution of \textit{interacting} binary systems\footnote{\footnotesize{By ``interacting'', we here refer to systems in which stars influence each other during their evolution.}}. However, it is possible to mimic the presence of unresolved binaries. The position of an unresolved binary depends on the mass of the primary and on the mass ratio. Once the mass of the primary star is set, the mass of the companion star is determined by a uniform random draw between 0.1 and 1 time the mass of the primary. The luminosity and the flux in the various filters of each component of the binary are summed. Because it is not easy to attribute a $T_\text{eff}$ to the pair, we assume that the $T_\text{eff}$ of the binary system is equivalent to that of the primary\footnote{\footnotesize{In the theoretical HRD, the luminosity (corrected for the binarity) is plotted versus the $T_\text{eff}$ (not corrected), while in the CMD, both colours and magnitudes are corrected.}}. 

Note that in case the mass of the secondary is lower than the minimal mass of the grid of stellar models used for the generation of the synthetic cluster, the star is tagged as a binary, but the secondary is not interpolated, and the fluxes are not corrected (only the flux of the primary is accounted for). In the HRD (or CMD), this produces a double tail near the lower limit of the IMF (one composed by the single stars and by the binaries for which only one component is accounted for, and the second composed by the binary with a mass-ratio close to one). As one moves towards higher mass, the gap between the two tails is progressively filled with binaries with various mass ratio. The fraction of unresolved binaries is an optional parameter of the code.

The photometric data of an observed cluster contains noise, which broadens the main sequence. In order to simulate the noise in the observed CMD, it is possible to add a gaussian noise (parametrised by a typical standard deviation $\sigma$ in the magnitude $M_\text{V}$ and the colour index B-V) to all the stars of the synthetic cluster.

\subsection{Evolution of star count ratios}

The aim of the ``star count'' mode is to predict the fraction of different types of stars as a function of the age given a fiducial (\textit{e.g.} a mass interval) for normalisation. The initial and final times between which the evolution is computed are an input for this option of the code. Then, a time step array is generated that ranges from the initial to the final time with a time step $\Delta t$. 

Another input is the number of mass ($m$) and rotation rate ($n$) cells to be considered. Then, we have $m+1$ values of initial mass, equally spaced in mass and $n+1$ values of initial rotation rates equally spaced as well.

The four vertex given by ($M_k,\omega_j$), ($M_k,\omega_{j+1}$), ($M_{k+1},\omega_j$), ($M_{k+1},\omega_{j+1}$), with $k$ between 1 and $m$ and $j$ between 1 and $n$ define a box. For each box, we assign the average value for the mass $M_{k+1/2}=(M_k+M_{k+1})/2$. The fraction $F_M$ of stars over the total population with a mass between $M_k$ and $M_{k+1}$ is given by the IMF, as described in a previous section. For this box of mass $M_{k+1/2}$ we have a fraction $F_{\omega}$ of stars with initial rotational rates between $\omega_j$ and $\omega_ {j+1}$, given by the assumed initial velocity distribution. In this way, the fraction of stars over the total population in each box at the ZAMS is given by $F_M \times F_{\omega}$.

Then, the evolution of each box is characterised by the evolution of a model of mass $M_{k+1/2}$, and its initial rotational rate $\omega_{j+1/2}$. At each time step, we count in the different boxes the number of stars of a given type, taking into account that stars of a certain mass disappear at that age (\textit{e.g.} explode as a supernova).

The quantities of interest for following throughout the evolution of a synthetic population include: the fraction of stars of a certain spectral type, such as B type stars (defined here as objects with effective temperatures between $10000\,\text{K}$ and $30000\,\text{K}$), the fraction of stars rotating above a certain rotation rate (\textit{e.g.} $\omega>0.8$), supergiant stars (with luminosities higher than $10^{4.9}L_\sun$), red supergiant stars \citep[RSG, with effective temperatures lower than $10^{3.66}\,\text{K}$) or the fraction of Be stars (which we consider to be dependent of the effective temperature, following][as explained in \citealt{Granada2013a}]{Cranmer2005a}.

\section{Results}\label{Sec_Results}

\subsection{On single stellar models}\label{Sec_StellarModels}
 
\begin{figure}
\centering
\hfill\includegraphics[width=0.47\textwidth]{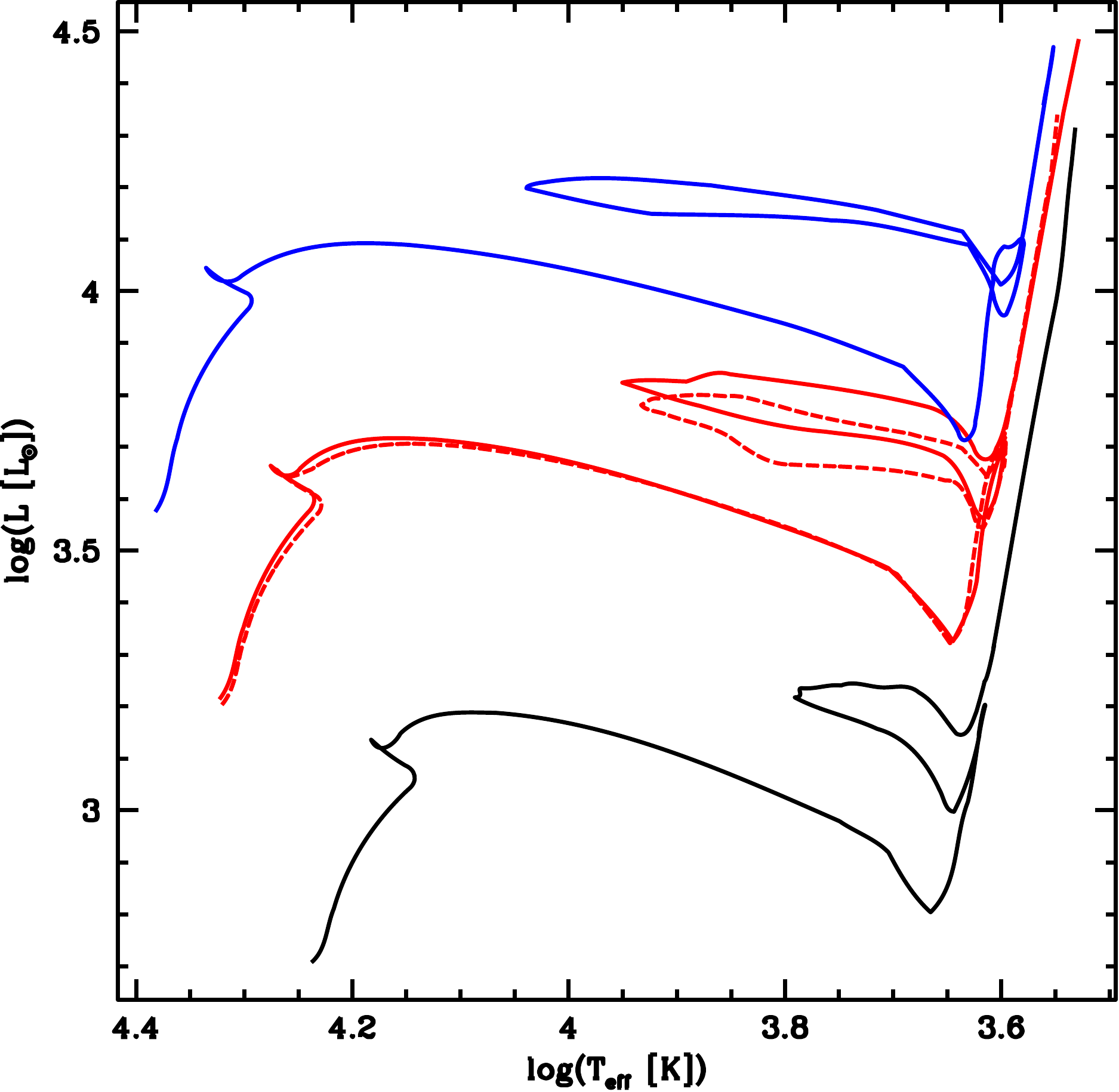}
\caption{
To illustrate the accuracy of the interpolation, we compare a $7\,M_\sun$ model with $\omega_\text{ini} = 0.5$ (red solid curve) obtained by interpolation between $5\,M_\sun$ (black) and $9\,M_\sun$ (blue) models of identical $\omega_\text{ini}$ with the ``real'' $7\,M_\sun$ model (red dashed curve) from \citet{Georgy2013a}.}
\label{Fig_Interpolation}
\end{figure}

We first check that interpolations of the tracks provide reasonable results by comparing tracks computed by the Geneva stellar evolution code with tracks obtained by interpolations of neighbouring tracks. An example of such an interpolation is shown on Fig.~\ref{Fig_Interpolation}, where we compare a $7\,M_\sun$ model at $Z=0.014$ with an initial rotation rate $\omega_\text{ini} = 0.5$ obtained by interpolation between a $5\,M_\sun$ and a $9\,M_\sun$ model with the same initial rotation rate, with a model that was rigorously computed with the Geneva stellar evolution code. We see that the MS and the HRD crossing, as well as the red supergiant (RSG) branch are well reproduced. For features depending on complex physical processes such as the blue loop, the interpolation provides less accurate results.

\begin{figure}
\centering
\includegraphics[width=0.47\textwidth]{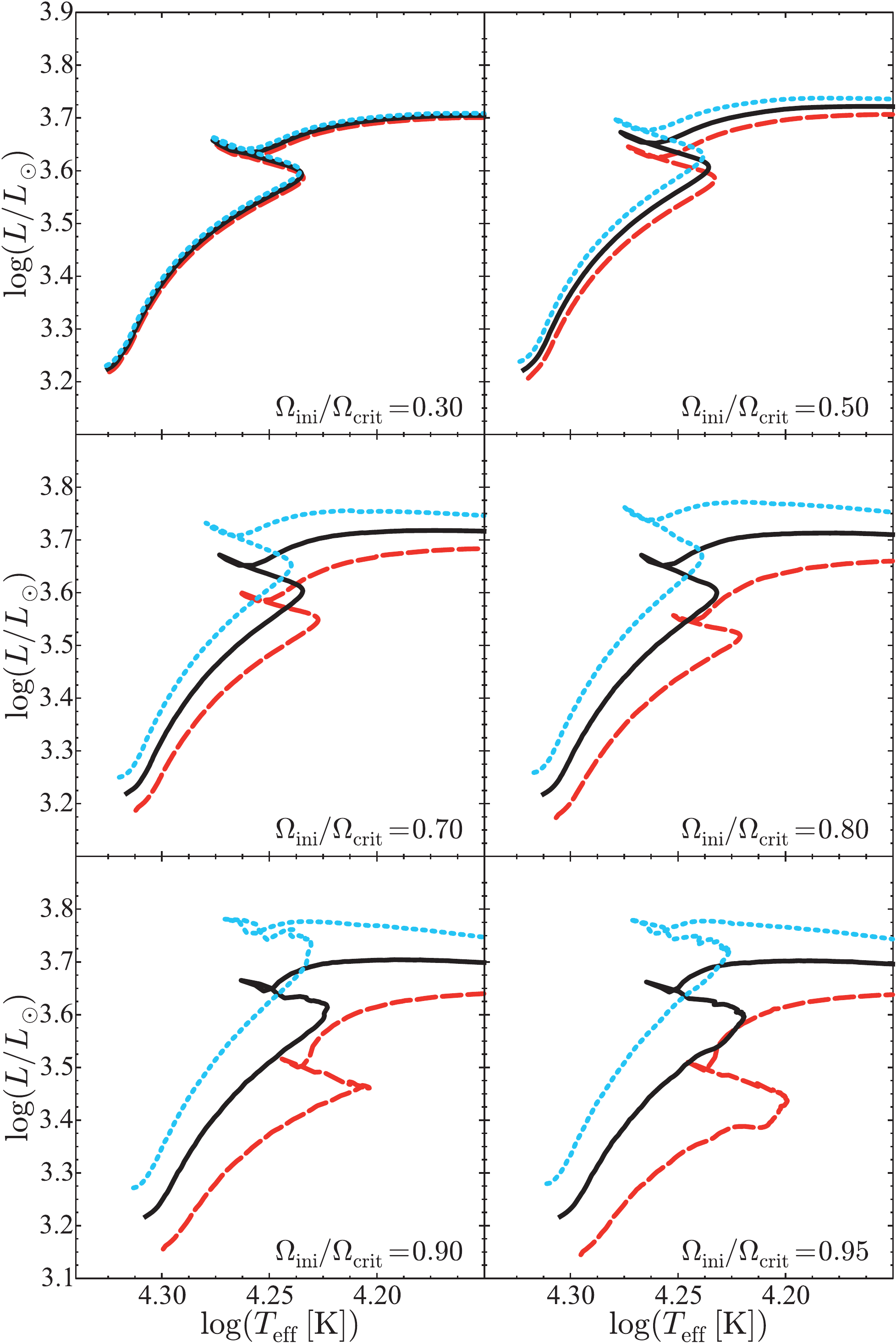}
\caption{Evolutionary tracks for $7\,M_\sun$ stellar models at $Z=0.014$ for various initial rotation rates, when the effects of gravity-darkening are accounted for. The blue dotted lines show the track for a star seen pole-on ($i=0^\circ$). The solid black and dashed red lines show the tracks with $i=45^\circ$, respectively $90^\circ$ (equator-on).}
\label{CompMS7}
\end{figure}

\begin{figure}
\centering
\includegraphics[width=0.47\textwidth]{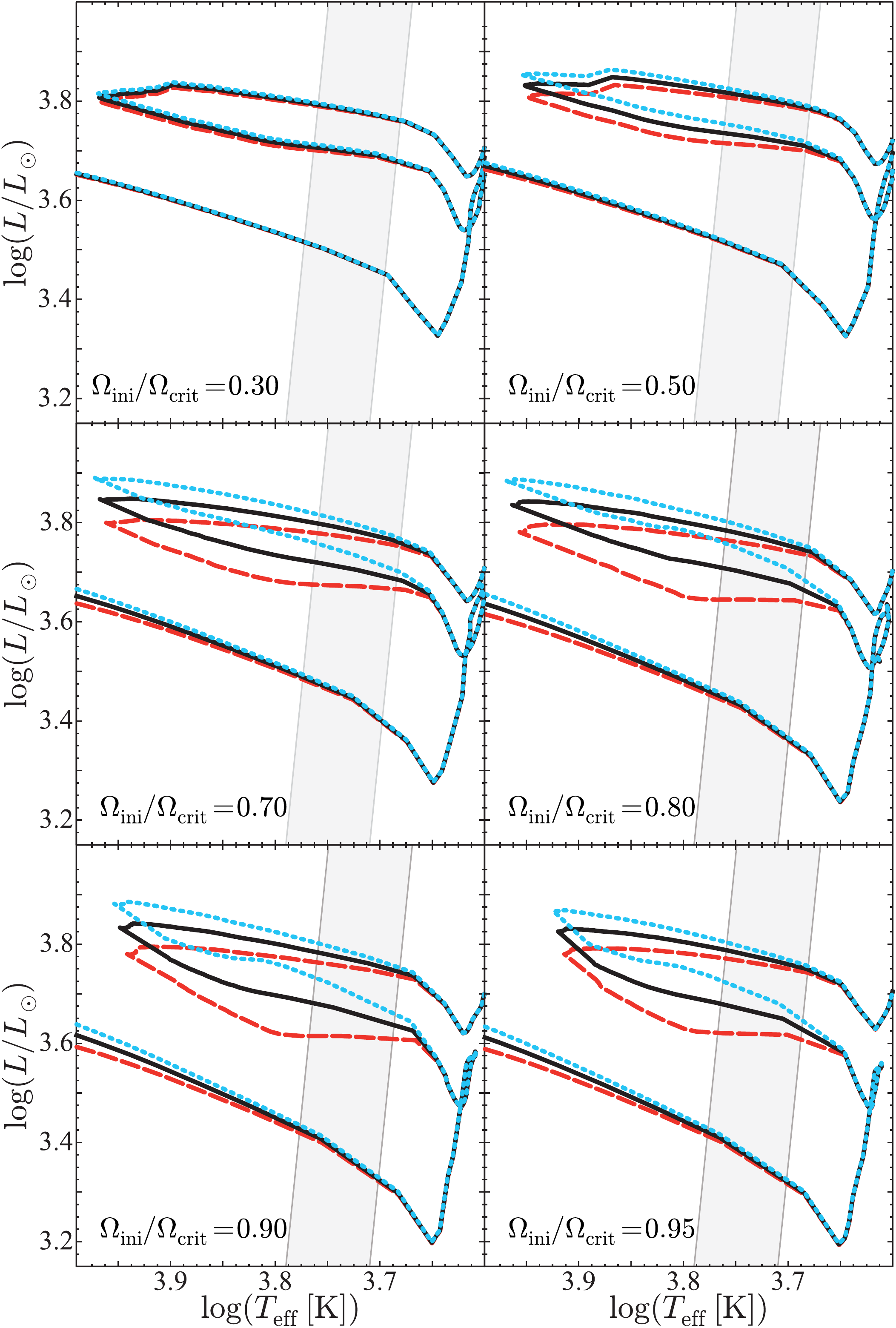}
\caption{Analogous to Fig.~\ref{CompMS7} for blue loops. The shaded area represents the Cepheid instability strip according to \citet{Tammann2003a}.}
\label{CompBL7}
\end{figure}

As explained in Sect.~\ref{Sec_GravDark}, tracks may be shifted in luminosity and effective temperature due to the gravity darkening effect. An illustration of this effect \citep[with the temperature law by][]{EspinosaLara2011a} is shown in Fig.~\ref{CompMS7}. As expected, tracks seen pole-on are more luminous and hotter than tracks seen equator-on (the effect of limb darkening is not accounted for here). The effect remains quite modest up to $\omega_\text{ini}=0.5$. Indeed the shift between the extreme points at the red turnoff at the end of the MS phase are at most of $0.005$ dex in effective temperature and $0.04$ dex in luminosity. At $\omega_\text{ini}=0.95$, the corresponding shifts are much larger: $0.027$ dex in $\log(T_\text{eff})$ and $0.28$ dex in $\log(L/L_\sun)$. 

In the plane $M_\text{V}$ versus B-V, these shifts are different. In this range of effective temperatures, B-V shows a very weak dependence on $\log(T_\text{eff})$ and thus the differences in effective temperature in the most extreme case ($\omega_\text{ini}=0.95$) will correspond to a change of only $0.002$ dex in B-V (for values of B-V around $-0.18$). The shift in $M_\text{V}$ is $0.105$ dex for $\omega_\text{ini}=0.50$ and $0.668$ dex for $\omega_\text{ini}=0.95$.

During the blue loops, the gravity darkening effect might also be noticeable, because the stars contract and their surface is thus accelerated, as illustrated in Fig.~\ref{CompBL7}. However $\omega$ does not reach extreme values and the shift remains lower than what is seen during the MS. Up to $\omega_\text{ini}=0.70$, the shift increases for increasing $\omega_\text{ini}$. For higher initial values, the maximal $\omega$ reached during the loop is similar, which explains the saturation of the gravity darkening effect that appears in Fig.~\ref{CompBL7} for the highest $\omega_\text{ini}$.

\begin{figure}
\centering
\includegraphics[width=.5\textwidth]{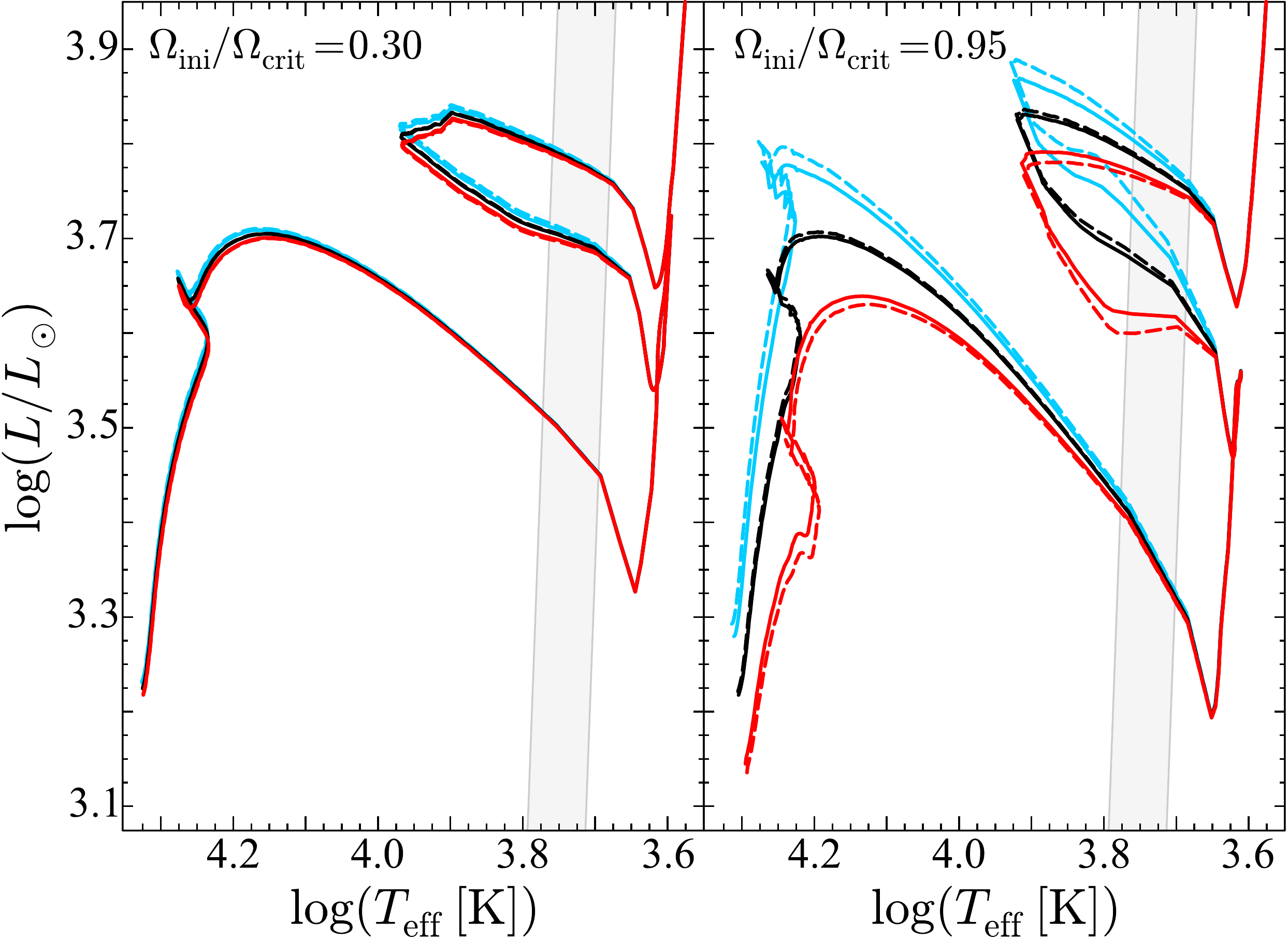}
\caption{Same as in Fig.~\ref{CompMS7}, with the addition of the effect of limb-darkening (dashed lines).}
\label{m7GDLD}
\end{figure}

In Fig.~\ref{m7GDLD}, the effect of limb-darkening has been added. It is barely visible on the MS for the relatively slowly rotating model shown in the left panel. It is more pronounced for the more rapid rotator shown in the right panel. The limb-darkening effect strengthens the effect of gravity darkening (see also Fig.~\ref{CORR}). Indeed, when the star is seen pole-on, the brightest regions of the star are seen perpendicularly, which tends to increase the observed flux. The opposite is true when the star is seen equator-on. 

\subsection{On isochrones}

Isochrone fitting is the most common technique of inferring the age of an observed cluster. It is thus interesting to study how the initial distribution of velocities among the cluster stars and the random orientation of the inclination angle impact the age determination via this method. We also study the way isochrones computed from a rotating stellar population differ with isochrones computed from non-rotating models.

\begin{figure}
\centering
\includegraphics[width=0.5 \textwidth]{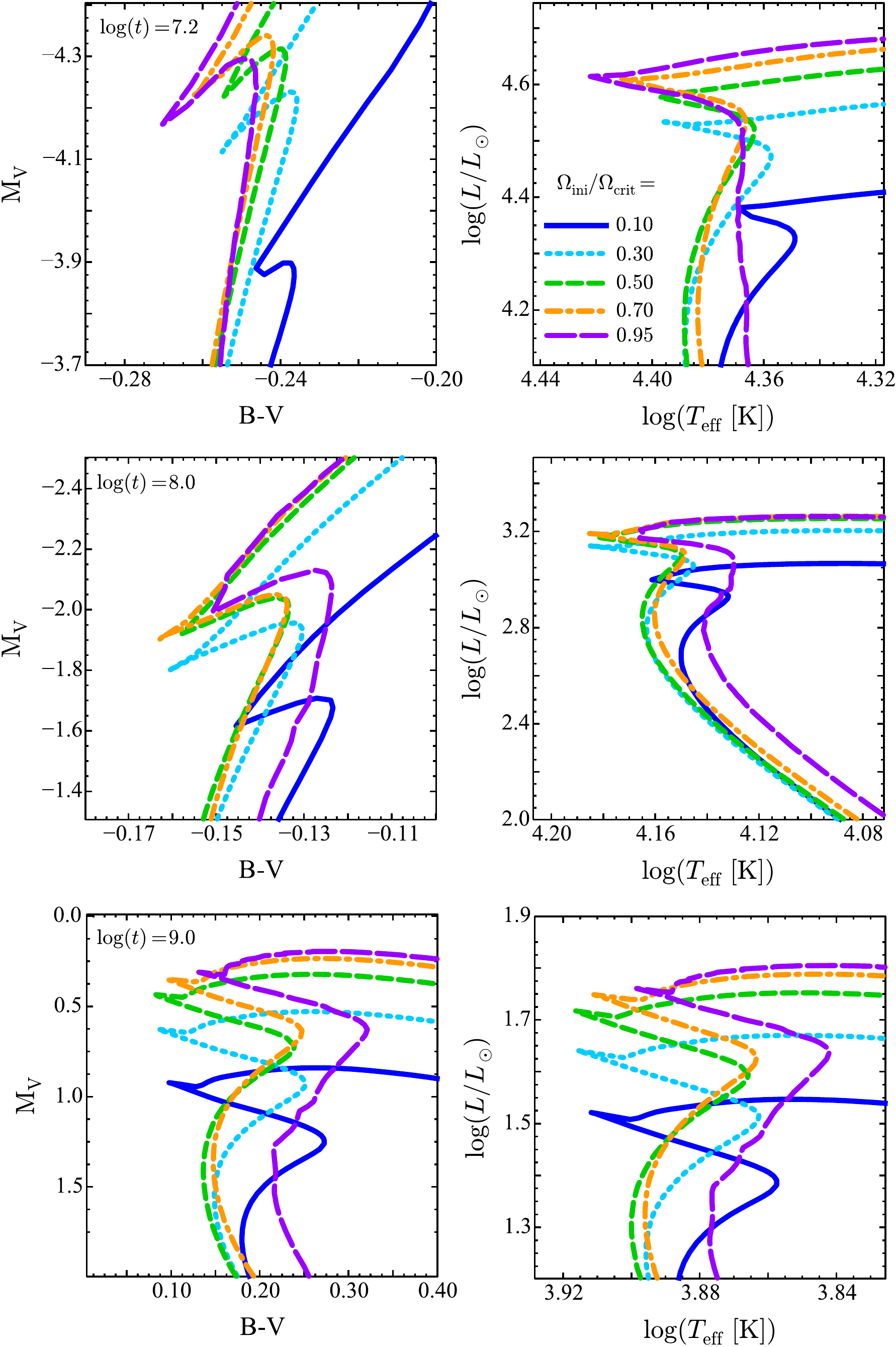}
\caption{Isochrones for various ages computed from stellar models with a Dirac initial velocity distribution. The initial velocities are indicated on the upper right panel. The mass at the turnoff for the logarithm of the  ages of $7.2$, $8.0$ and $9.0$ are around $12$, $5$ and $2\,M_\sun$, respectively.}
\label{Compis}
\end{figure}

In Fig.~\ref{Compis}, we show isochrones in the $M_\text{V}$ versus B-V plane and  in the $\log(L/L_\odot)$ versus $\log(T_\text{eff})$ plane. These isochrones are computed assuming stars with an identical $\omega_\text{ini}$. This is of course not realistic, although the isochrones corresponding to $\omega_\text{ini}=0.3$ and $0.5$ should encompass most of the observed stars. Note that in ``Isochrone'' mode, the effects of gravity darkening are not taken into account.

The isochrones corresponding to initially fast rotating stars have turnoff at a higher luminosity and in general at a bluer colours than isochrones computed from slowly or non-rotating  models. Comparing $\text{M}_\text{V}$ at the turnoff for the isochrones with $\omega_\text{ini}=0.1$ and $0.7$,  the difference amounts to $0.4$ dex at $\log(t) = 7.2$ and to $0.55$ dex at $\log(t) = 9.0$.  At very high rotation ($\omega_\text{ini}=0.95$), the tracks may be shifted to the right (in cooler regions) because hydrostatic effects become dominant with respect to the mixing effect in the mass range considered.

\subsection{On synthetic stellar clusters}

In contrast to isochrones, synthetic CMDs account for the fact that a stellar population is not only composed of stars of various masses but also of stars with different initial rotation velocities and angles of view.

In this section, we use our population synthesis code as an instrument to analyse single-aged stellar populations from a purely theoretical point of view. The main questions we intend to address are the following: how does the velocity distribution evolve as a function of time? How does the inclusion of rotation impact the determination of the ages and of the mass at the turnoff? What is the impact of different distributions of initial rotation? How do gravity and limb-darkening distort the CMDs? Are the initially rapid rotators located in a peculiar place in the cluster CMD? Where are the most rapid rotators? Where are the nitrogen-enriched stars? What are the effects of stochasticity on the appearance of a given CMD cluster? How do the above properties change when the age increases?

\subsubsection{Evolution of the velocity distribution}

In this section, we study how the distribution of rotational velocities varies in clusters of different ages. To reduce the effects of stochasticity, we consider a population initially composed of $10000$ stars, distributed within the mass range $1.7$ to $15\,M_\sun$ according to a Salpeter's IMF.

\begin{figure*}
\centering
\includegraphics[width=\textwidth]{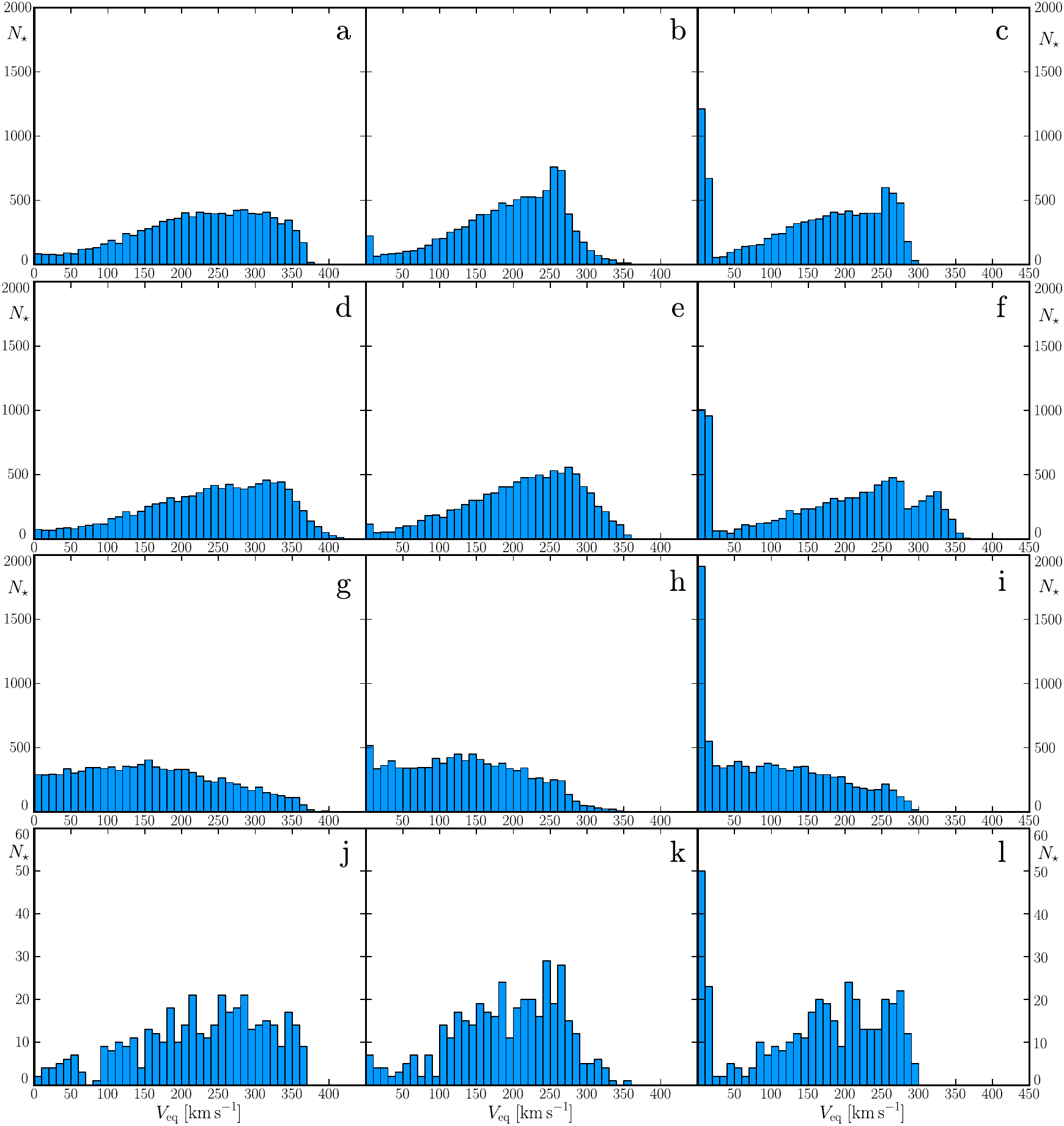}
\caption{Distribution of surface equatorial velocities in a cluster at three different ages (first column: $10\,\text{Myr}$, second column: $100\,\text{Myr}$, and third column: $1\,\text{Gyr}$). \textit{Top row (panels a, b, c):} $Z_\sun$ cluster with the initial velocity distribution from \citet[][see their Fig.~6]{Huang2010a}. \textit{Second row (panels d, e, f):} same as above, but at $Z=0.002$. \textit{Third row (panels g, h, i):} same as top row, but with the initial velocity distribution from \citet{Huang2006a}. \textit{Bottom row (panels j, k, l):} same as top row but for a total number of stars of $400$ instead of $10000$.}
\label{Vel_distr}
\end{figure*}

The first row of Fig.~\ref{Vel_distr} (panels a, b, c) shows the time evolution of the distribution of rotational velocities between $10\,\text{Myr}$ and $1\,\text{Gyr}$ when the initial rotation velocity distribution by \citet{Huang2010a} is used at solar metallicity ($Z=0.014$). In panel a, the initial masses of stars span the whole mass interval between $1.7$ and $15\,M_\sun$. In panels b and c, the mass interval is reduced to respectively $[1.7, 5.7]$ and $[1.7,2.4]$, due to the fact that more massive stars progressively disappear. Between $10$ and $100$ Myr, the maximum velocity of the distribution decreases and the peak of the distribution shifts toward a lower velocity. This is a quite general trend that comes from the decrease of the rotation rate occuring at the very beginning of the evolution (see for instance Fig.~6 in \citealt{Georgy2013a} or \citealt{Granada2013a}). This effect depopulates the high velocity end of the distribution ($V_\text{eq}\gtrsim300\ \text{km\,s}^{-1}$). At $100$ Myr, this braking has affected almost the whole mass range of the cluster. In addition, the strong braking that occurs after the end of the MS when the star evolves to become a red (super)giant progressively populates the very low velocity end of the distribution.

The second row of Fig.~\ref{Vel_distr} (panels d, e, f) shows the same velocity distribution evolution at a lower metallicity $Z=0.002$. The global trend is similar, except that there are generally more rapid rotators. Indeed the \citet{Huang2010a} initial velocity distribution is given in terms of $V_\text{ini}/V_\text{crit}$ which corresponds to slightly higher velocities when the metallicity decreases (the stars are more compact at low $Z$). In the 1 Gyr panel (f), the distribution becomes double-peaked. Two causes are identified: first, at this age, only the lowest mass stars are still on the MS, and in this mass range $V_\text{eq}$ increases during the MS (in contrast with the $Z_\sun$ case). Second, rapid rotation significantly increases the MS duration in this mass domain, and for a given mass only the most rapid rotators are still on the MS.

The third row of Fig.~\ref{Vel_distr} (panels g, h, i) shows the results obtained when the velocity distribution of \citet{Huang2006a} is used as the initial one. This distribution favours the moderate rotators. In particular, the peak of the velocity distribution at an age of $10\,\text{Myr}$ is shifted from a value of approximately $280\,\text{km\,s}^{-1}$ to a value of $150\,\text{km\,s}^{-1}$ and the overall distribution is flatter than the one obtained with the \citet{Huang2010a} initial velocity distribution.

The bottom row (panels j, k, l) shows the same distribution as shown in panels a, b, and c, albeit for a cluster with an initial number of stars of 400. This illustrates the impact of stochastic effects on the distribution, which is to produce fluctuations. However the general trends are preserved, and show that a few hundred stars provide a sufficient sampling for obtaining a good description of the velocity distribution.

\subsubsection{Synthetic colour-magnitude diagram for a 100 Myr cluster}

\begin{table}[t]
\caption{Some properties of the stars labelled in Fig.~\ref{CMD1}. The alphabetical labels are for MS stars, and the numerical ones for evolved stars. The stars are sorted by increasing mass.}
\begin{center}
\scalebox{0.92}{
\begin{tabular}{cccrrrr}
\hline
\rule[0mm]{0mm}{5mm}Label & $M$ & $\omega_\text{ini}$ & \multicolumn{1}{c}{$V_\text{eq}$} &  \multicolumn{1}{c}{$\frac{^{12}\text{C}/^{13}\text{C}}{\left(^{12}\text{C}/^{13}\text{C}\right)_\text{ini}}$} &  \multicolumn{1}{c}{$\frac{\text{N}/\text{C}}{\left(\text{N}/\text{C}\right)_\text{ini}}$} &  \multicolumn{1}{c}{$\frac{\text{N}/\text{O}}{\left(\text{N}/\text{O}\right)_\text{ini}}$}  \\
\rule[-3mm]{0mm}{5mm} & $[M_\sun]$ & & $[\text{km s}^{-1}]$ & & &\\
\hline
A	&	$4.26$	&	$0.06$	&	$13.$	&	$1.0$	&	$1.0$	&	$1.0$\rule[0mm]{0mm}{3mm}\\
B	&	$4.38$	&	$0.96$	&	$322.$	&	$0.4$	&	$1.7$	&	$1.5$\\
C	&	$4.44$	&	$0.69$	&	$199.$	&	$0.5$	&	$1.4$	&	$1.3$\\
D	&	$4.52$	&	$0.66$	&	$188.$	&	$0.5$	&	$1.5$	&	$1.4$\\
E	&	$4.62$	&	$0.85$	&	$265.$	&	$0.3$	&	$2.3$	&	$1.9$\\
F	&	$4.73$	&	$0.39$	&	$101.$	&	$0.8$	&	$1.1$	&	$1.1$\\
G	&	$4.75$	&	$0.52$	&	$141.$	&	$0.6$	&	$1.3$	&	$1.2$\\
H	&	$4.77$	&	$0.98$	&	$351.$	&	$0.3$	&	$2.6$	&	$2.0$\\
I	&	$4.82$	&	$0.23$	&	$53.$	&	$1.0$	&	$1.0$	&	$1.0$\\
J	&	$4.83$	&	$0.72$	&	$210.$	&	$0.3$	&	$2.4$	&	$1.9$\\
K	&	$4.99$	&	$0.73$	&	$209.$	&	$0.3$	&	$2.9$	&	$2.2$\\
L	&	$5.03$	&	$0.86$	&	$272.$	&	$0.3$	&	$3.6$	&	$2.6$\\
M	&	$5.04$	&	$0.74$	&	$214.$	&	$0.3$	&	$3.1$	&	$2.4$\\
N	&	$5.05$	&	$0.92$	&	$307.$	&	$0.3$	&	$3.8$	&	$2.7$\\
O	&	$5.10$	&	$0.44$	&	$108.$	&	$0.5$	&	$1.5$	&	$1.4$\\
P	&	$5.24$	&	$0.94$	&	$324.$	&	$0.2$	&	$6.4$	&	$3.6$\\
1	&	$5.16$	&	$0.14$	&	$9.$		&	$0.2$	&	$6.2$	&	$4.0$\\
2	&	$5.18$	&	$0.30$	&	$2.$		&	$0.2$	&	$8.5$	&	$5.0$\\
3	&	$5.22$	&	$0.49$	&	$37.$	&	$0.4$	&	$1.8$	&	$1.6$\\
4	&	$5.24$	&	$0.58$	&	$57.$	&	$0.3$	&	$2.5$	&	$2.0$\\
5	&	$5.35$	&	$0.68$	&	$5.$		&	$0.1$	&	$16.9$	&	$6.5$\\
6	&	$5.38$	&	$0.82$	&	$6.$		&	$0.1$	&	$22.7$	&	$7.0$\\
7	&	$5.41$	&	$0.78$	&	$7.$		&	$0.1$	&	$21.2$	&	$6.9$\\
8	&	$5.48$	&	$0.51$	&	$13.$	&	$0.2$	&	$12.5$	&	$5.9$\\
9	&	$5.51$	&	$0.59$	&	$15.$	&	$0.1$	&	$14.8$	&	$6.3$\\
\hline
\end{tabular}}
\end{center}
\label{tabla0}
\end{table}

We first compute a synthetic cluster with $400$ stars in the mass range between $1.7$ and $15\,M_\sun$ and an age of $100\,\text{Myr}$ at $Z_\sun$. The initial velocity distribution is the one from \citet{Huang2010a}, and the inclination angles follow a random distribution. The cluster contains $30\%$ of unresolved binaries, following the result by \citet{Oudmaijer2010a}\footnote{\footnotesize{We assume that all binaries will appear as unresolved.}}. A gaussian noise in the visible magnitudes and colours with standard deviation of $0.15\,\text{mag}$ and $0.10\,\text{mag}$, respectively, has been added. We used the calibrations between colour, bolometric corrections and effective temperatures given by \citet{Worthey2011a}.

\begin{figure}
\centering
\includegraphics[width=0.5\textwidth]{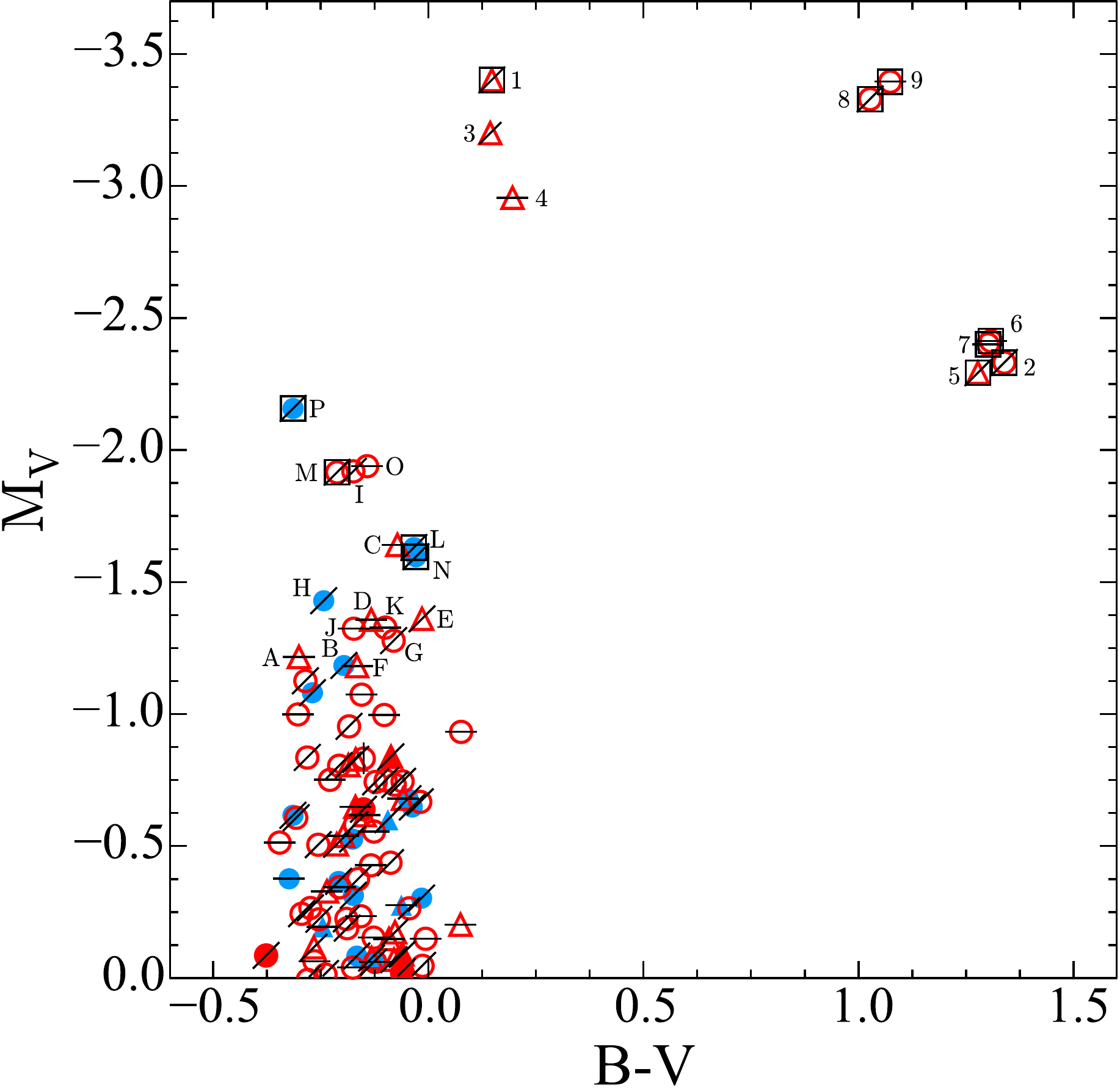}
\caption{Synthetic CMD for a $100\,\text{Myr}$ old cluster at $Z_\sun$, computed with the initial velocity distribution of \citet{Huang2010a}. A Gaussian noise with a standard deviation of $0.15\,\text{mag}$ in $M_\text{V}$ and $0.1\,\text{mag}$ in B-V has been added. Different characteristics are indicated by different symbols. \textit{Multiplicity:} single stars (circles) or unresolved binaries (triangles); \textit{velocity:} current $\omega<0.85$ (red), or $\omega\geq0.85$ (blue), with filled symbols for $\omega_\text{ini}\geq0.85$; \textit{inclination angle:} $i\geq70^\circ$ (horizontal stroke), $i<10^\circ$ (vertical stroke), intermediate inclinations (diagonal stroke); \textit{surface chemical enrichment:} $\text{(N/C)/(N/C)}_\text{ini}\geq3$ (surrounding square). The labelled stars are listed in Table~\ref{tabla0}.}
\label{CMD1}
\end{figure}

Figure~\ref{CMD1} shows the CMD of this cluster. Fast rotators are more frequent towards the turnoff of the cluster sequence than below, along the MS band. Typically, in the magnitude range between $-1$ and $-2$, the total number of stars is $18$. Among these, $5$ stars have an $\omega >0.85$ ($28\%$ of the sample). Between the magnitudes $0$ and $-1$, there are $13$ fast rotators among a total of $69$ stars ($18\%$). Indeed, rotation enhances the MS lifetime of a given initial mass star and increases its luminosity. This favours fast rotators in the upper part of the MS band.

Among the $18$ fast rotators present in the whole CMD, all began their evolution with an initial fast rotation ($\omega_\text{ini}>0.85$).

The random angle distribution implies that very few stars will be seen pole-on. Actually, between the magnitudes $0$ and $-2$, only $1$ star is seen nearly pole-on ($i<10^\circ$) and $33$ stars are seen equator-on ($i\geq70^\circ$) among a total of $87$ stars.

All significantly nitrogen-enriched stars are either at the turnoff or evolved post-MS stars. The MS stars with a N/C greater than $3$ times the initial ratio span the range of values between $3.1$ (star M)\footnote{\footnotesize{The labels shown here can be retrieved in Table~\ref{tabla0} and Fig.~\ref{CMD1}.}} and $6.4$ (star P). Among the post-MS stars, the N/C enhancement factor lies between $1.8$ (star 3) and $22.7$ (star 6). Stars 2 and 5 lie nearly in the same position of the HRD, although their N/C ratios differ greatly. The enrichment in star 2 (8.5) is mainly due to the dredge-up process with only a small contribution from the rotational mixing, since it has $\omega_\text{ini}=0.30$. The star 5, on the other hand, has $\omega_\text{ini}=0.68$, which explains the much higher enhancement factor (16.9). There is no clear correlation between the actual surface velocity and the nitrogen enrichment. Other effects like the initial mass, and hence the evolutionary phase, produce some dispersion. The behaviour of the N/O and $^{12}\text{C}/^{13}\text{C}$ is generally the same as N/C. This is due to the fact that both of these ratio are affected by H burning via the CNO cycle, and thus follow the same trend as the N/C ratio.

Due to rotation, there is some overlap between the mass range of stars in the upper part of the MS band and the mass range of post-MS stars. The minimum initial mass of the post-MS stars is $5.16\,M_\sun$ (star 1) and its maximum value is $5.51\,M_\sun$ (star 9). On the MS band, the maximum mass is $5.24\,M_\sun$ (star P). Thus the mass range between $5.16$ and $5.24\,M_\sun$ can be found both on the MS band and in the post-MS phases, depending on initial rotation. For instance, star P, which is still on the MS band, was initially a rapid rotator ($\omega_\text{ini}=0.94$), while star 2, which is an evolved star with roughly the same initial mass, was initially a slowly rotating star ($\omega_\text{ini}=0.30$).

Were this synthetic cluster an observed one, how would the age determination by the isochrone fitting method be impacted by not considering some of the effects mentioned in Sect.~\ref{SectSynthClust}?

\begin{figure*}
\centering
\includegraphics[width=0.48\textwidth]{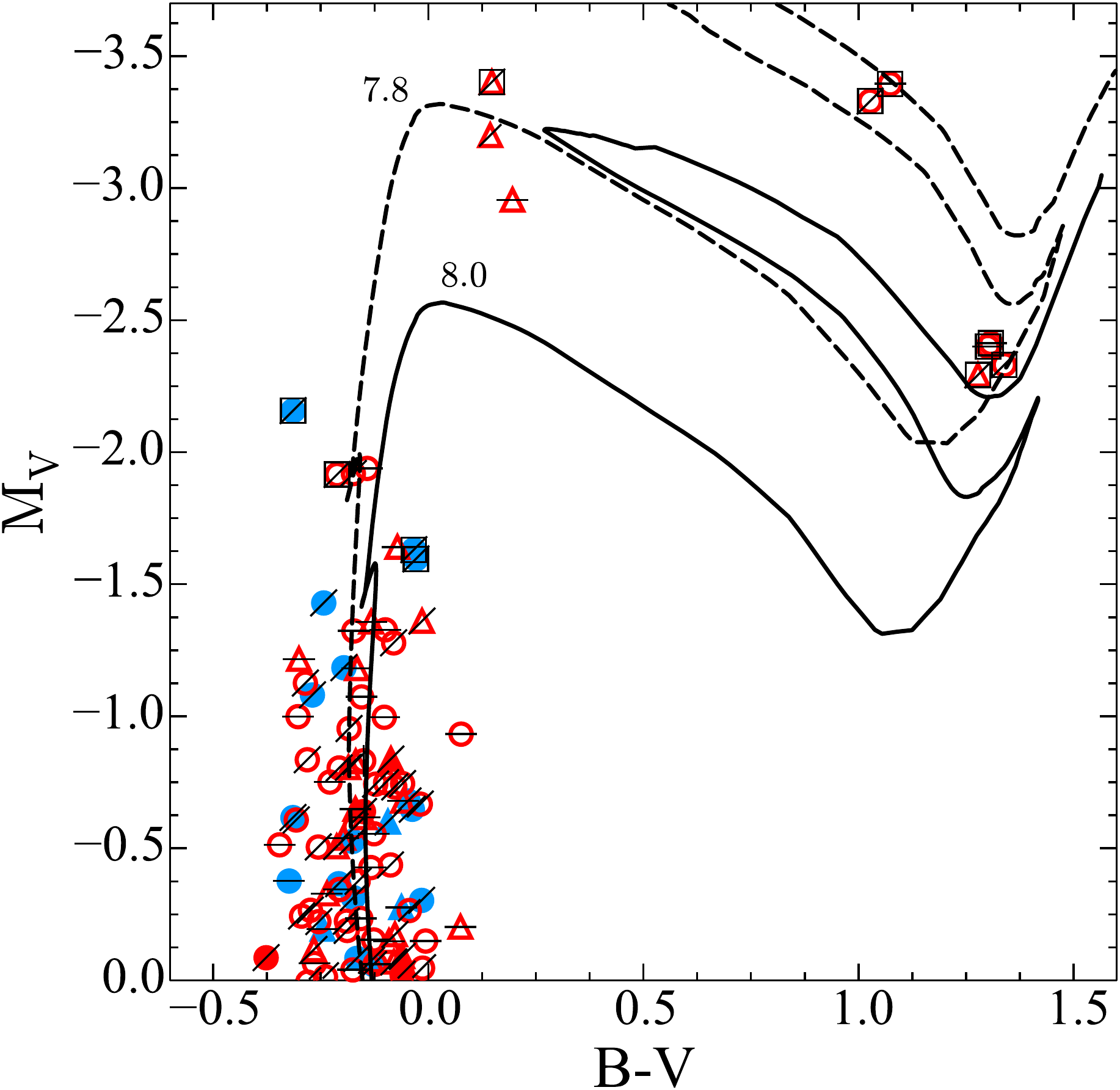}\hfill\includegraphics[width=0.48\textwidth]{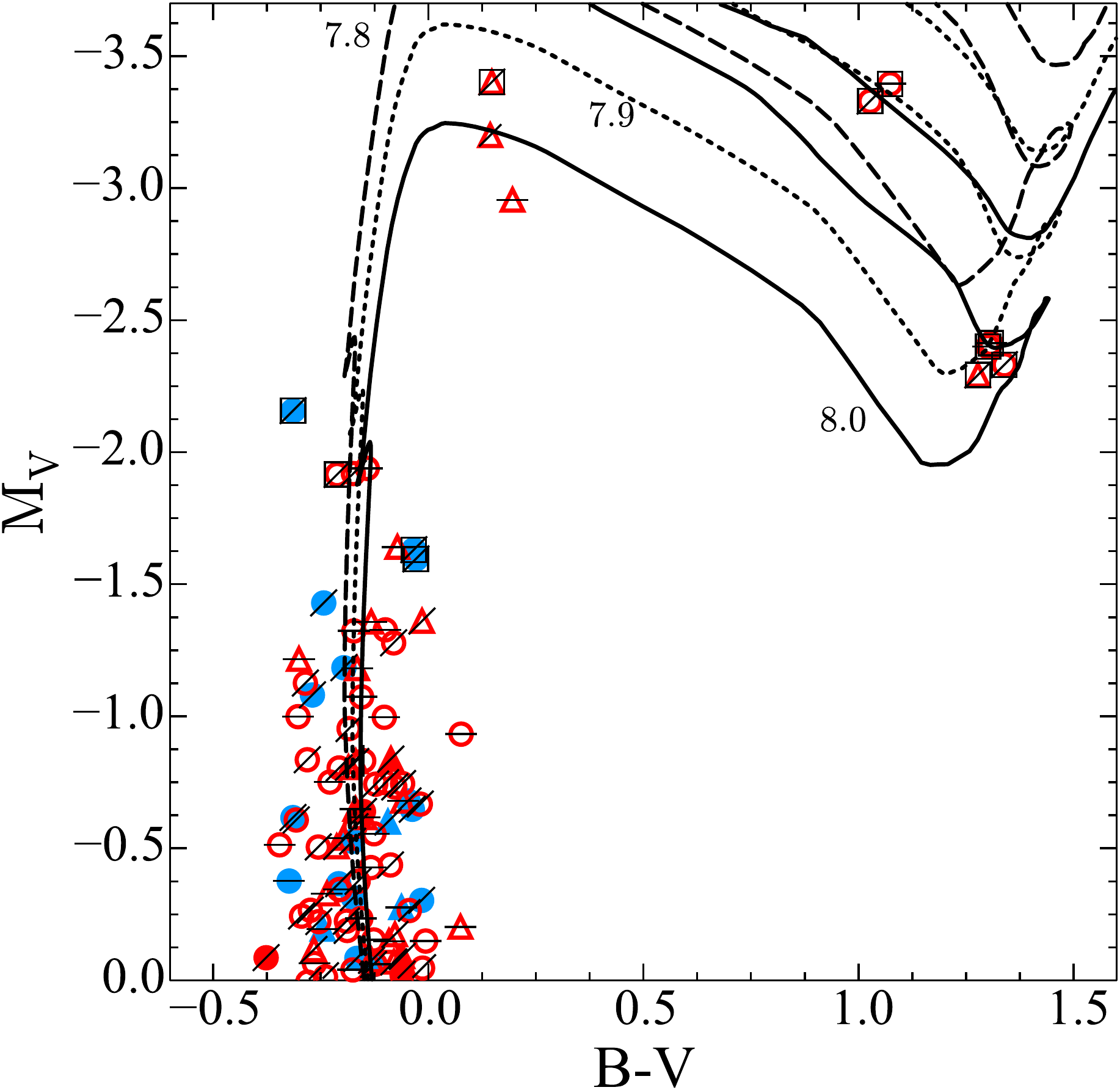} 
\caption{Same cluster as Fig.~\ref{CMD1} with isochrones superimposed. \textit{Left panel:} isochrones computed from non-rotating models. \textit{Right panel:} isochrones computed from rotating models with $V_\text{ini}/V_\text{crit}=0.4$.
}
\label{CMDV}
\end{figure*}

In Fig.~\ref{CMDV} (\textit{left panel}), isochrones obtained from non-rotating models are superimposed to the same cluster as in Fig.~\ref{CMD1}. The isochrone corresponding to the actual age of the cluster does not fit the magnitude of the turnoff. An isochrone at younger age ($\log(t) \lesssim 7.8$) would provide a better fit. However, the inferred age of the cluster would be younger by $40\%$ than the true one. Moreover the mass at the turnoff for the $V_\text{ini}=0$ isochrone at $\log(t)=7.8$ is $5.8\,M_\sun$, while the actual masses of stars at the turnoff (I, M, O, P) are between $4.8$ and  $5.2\,M_\sun$.

\begin{figure}
\centering
\includegraphics[width=0.5\textwidth]{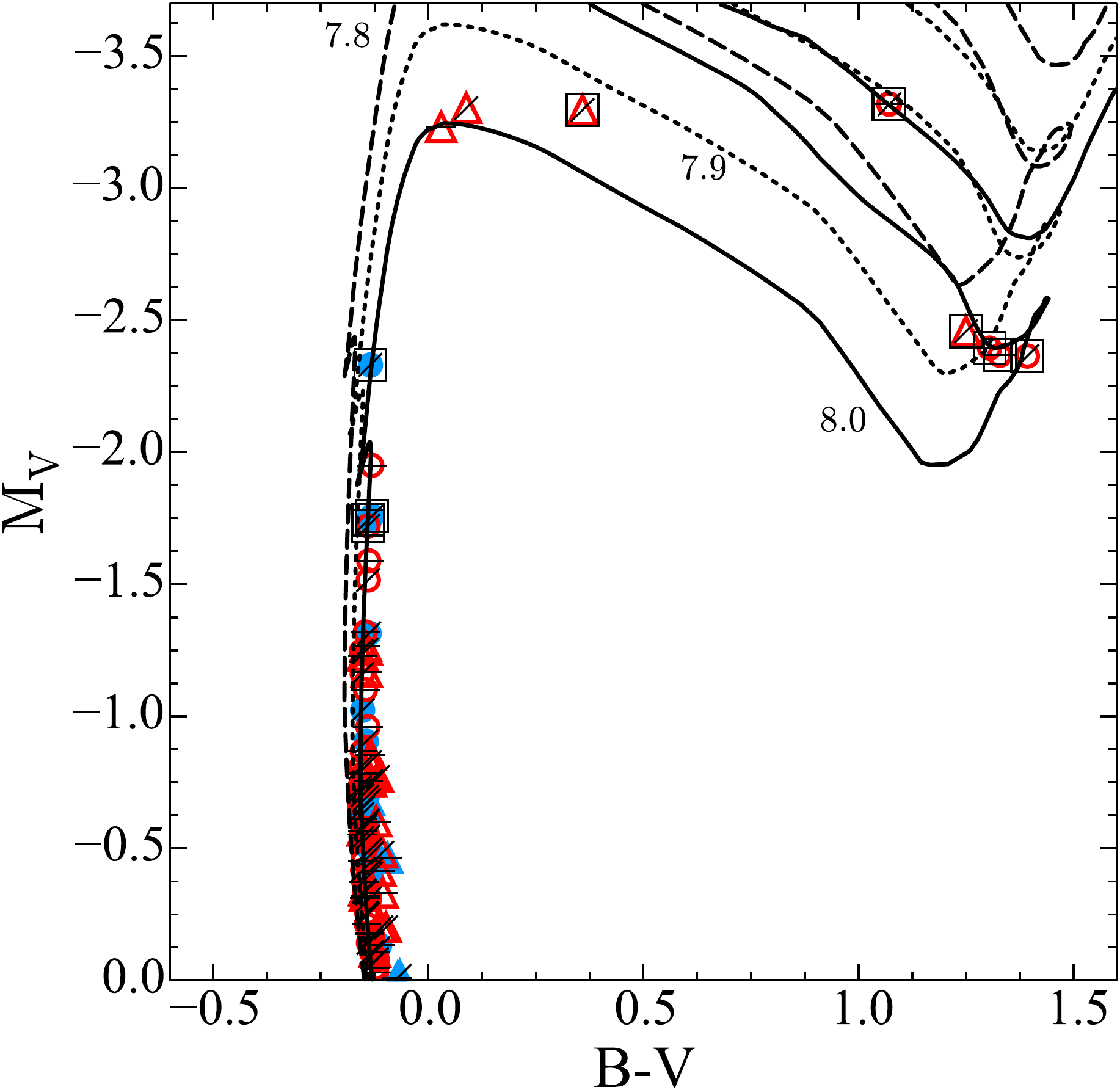}
\caption{Same synthetic cluster as the one shown in Fig.~\ref{CMD1} but where the photometric errors have been removed.}
\label{CMDN}
\end{figure}

\begin{figure}
\centering
\includegraphics[width=0.24\textwidth]{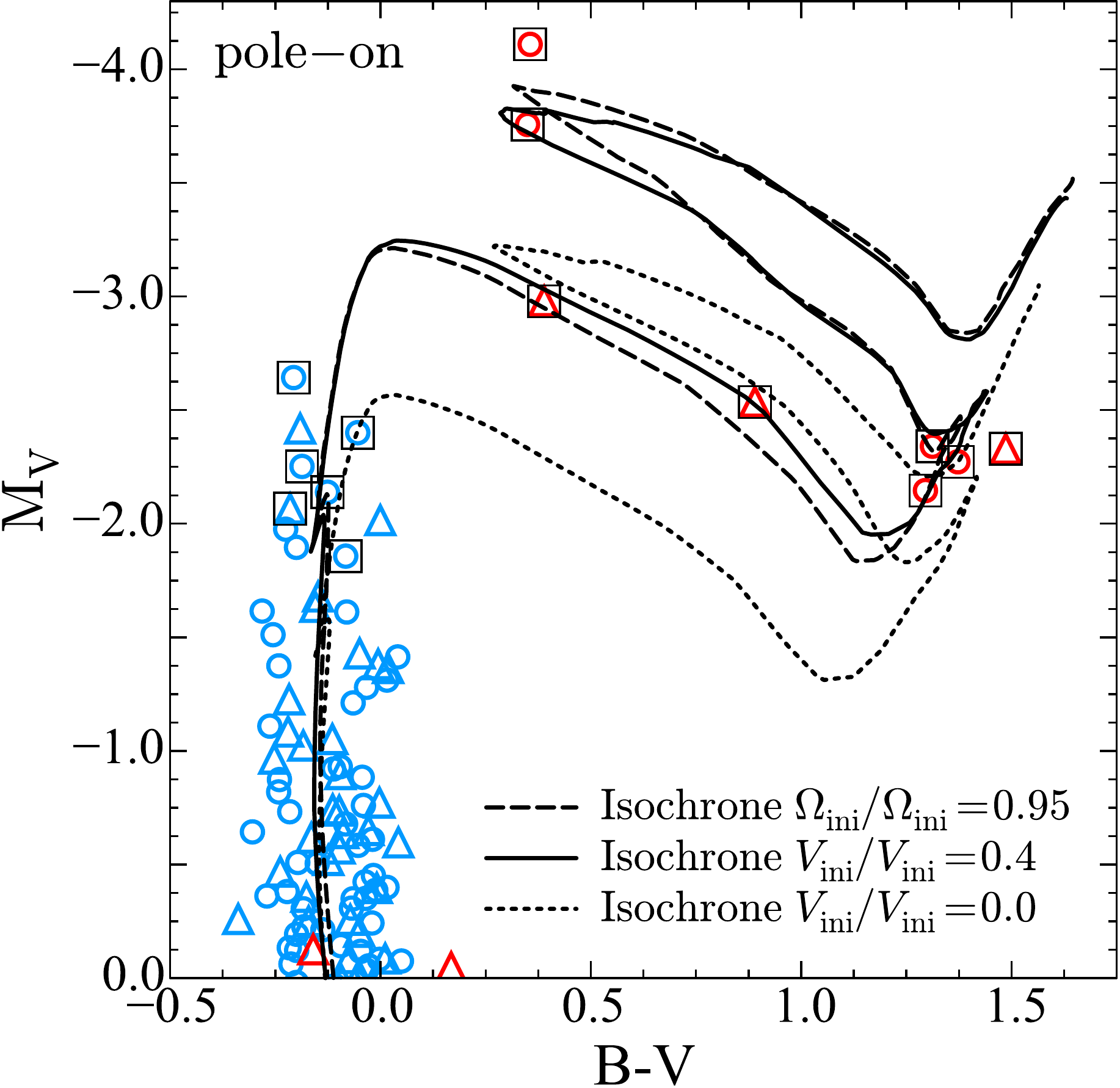}
\includegraphics[width=0.24\textwidth]{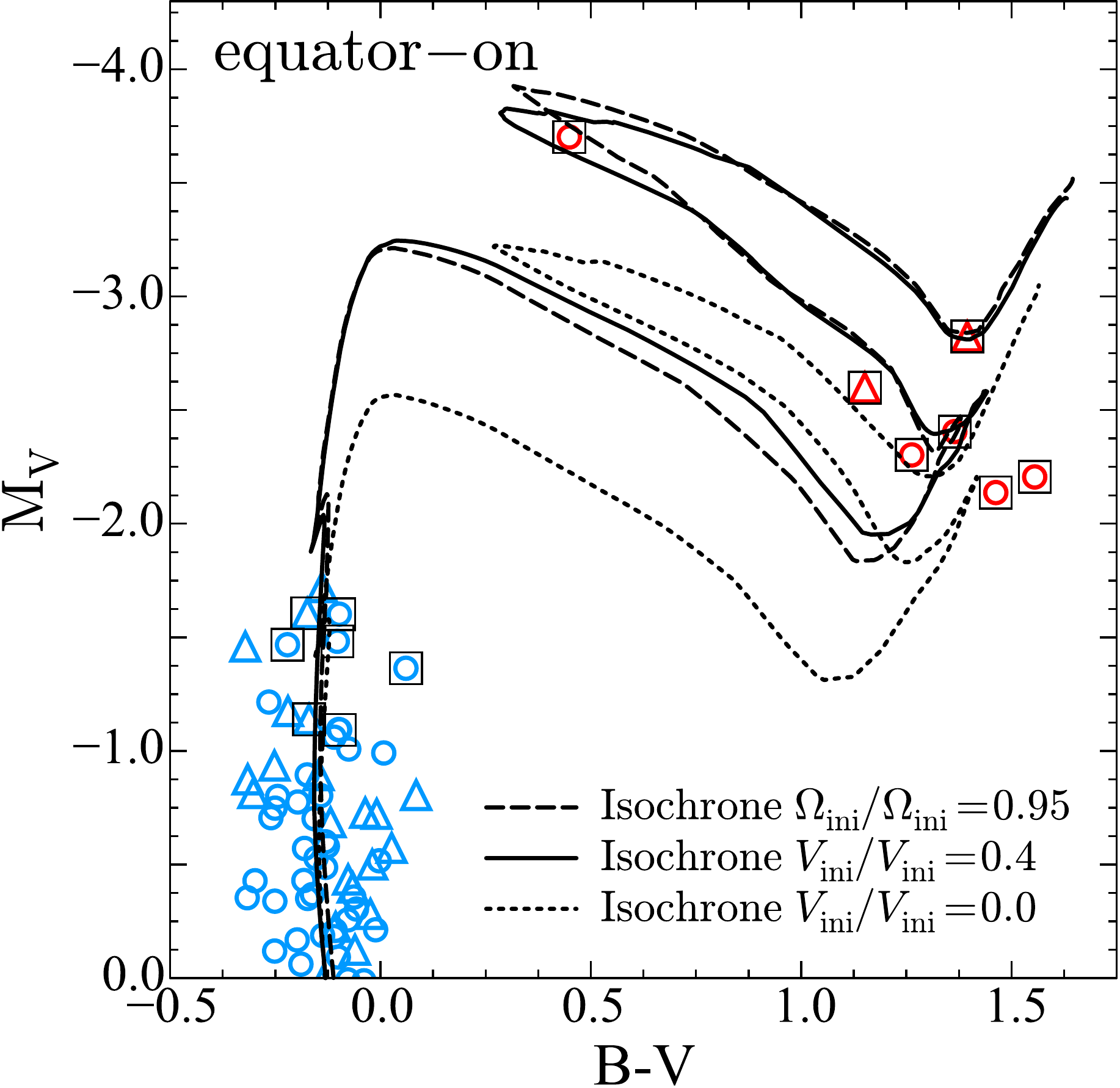}
\caption{Synthetic cluster with similar characteristics as the one shown in Fig.~\ref{CMD1}, but where all the stars began their evolution with $\omega_\text{ini}=0.95$. Isochrones for various initial rotation velocities and for an age of $\log(t) = 8$ are also shown. \textit{Left panel:}  all stars are seen pole-on. \textit{Right panel:}  all stars are seen equator-on.}
\label{CMDA}
\end{figure}

\begin{figure}
\centering
\includegraphics[width=0.5\textwidth]{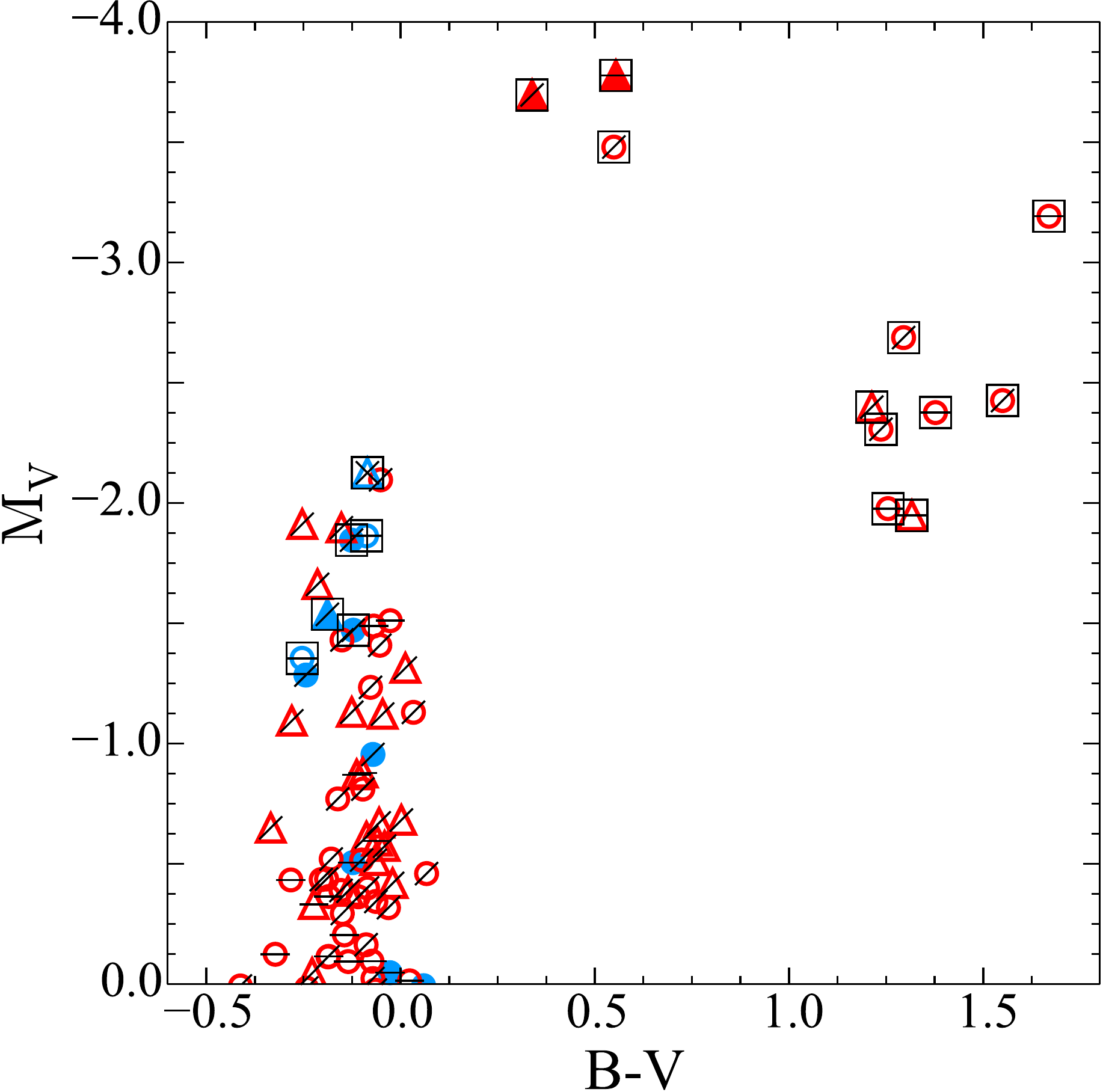}
\caption{Same synthetic cluster as the one shown in Fig.~\ref{CMD1} but with a different initial distribution of the velocities (see text).}
\label{CMDH}
\end{figure}

When isochrones obtained from rotating models are used (Fig.~\ref{CMDV}, \textit{right panel}), the situation is improved. However, the turnoff magnitude of the correct isochrone ($\log(t)=8.0$) is still fainter than the more luminous MS star of the cluster, and the best fit is provided by the $\log(t)=7.9$ isochrone, which underestimates the true age of the cluster by $20\%$. Figure~\ref{CMDN}, which presents the same cluster without photometric noise, shows that the age discrepancy remains and is thus mostly due to the rotational mixing acting inside star P.

Note that the isochrones were built from models with only one $V_\text{ini}/V_\text{crit}$ (the average rotation rate $V_\text{ini}/V_\text{crit}=0.4$). The extreme stars are not expected to be reproduced. However, as our grids of rotating stellar models were calibrated in order to match the observed features at solar metallicity \citep[see][]{Ekstrom2012a}, we expect the isochrones built from these models to give the best account for the clusters.

Figure~\ref{CMDA} shows two clusters with the same characteristics as the one presented in Fig.~\ref{CMD1}, but where all stars are initially rapid rotators ($\omega_\text{ini}=0.95$) and are seen either pole-on (\textit{left panel}) or equator-on (\textit{right panel}), so that the impact of gravity- and limb-darkening is maximised. As expected, the turnoff is brighter by about $0.9$ dex in M$_\text{V}$ in the pole-on cluster than in the equator-on one. Since these synthetic CMDs result from two different draws, stochastic effects produce some scatter, but the general trend is clear. Of course, this is an upper limit value due to the very extreme cases considered here. In B-V, there is also a shift, but it is so small that it is drowned in the photometric noise.

The velocity distribution of \citet{Huang2010a} has been obtained from young stars, and thus is convenient to be used as an initial velocity distribution in SYCLIST. \citet{Huang2006a} have also proposed a $V_\text{eq}$ distribution\footnote{\footnotesize{The conversion from the original $V_\text{eq}$ distribution to a $\omega$ distribution is done in \citet{Ekstrom2008b}}}, and it is interesting to weight the impact of the use of a different distribution (Fig.~\ref{CMDH}).

\begin{figure*}
\centering
\includegraphics[width=.8\textwidth]{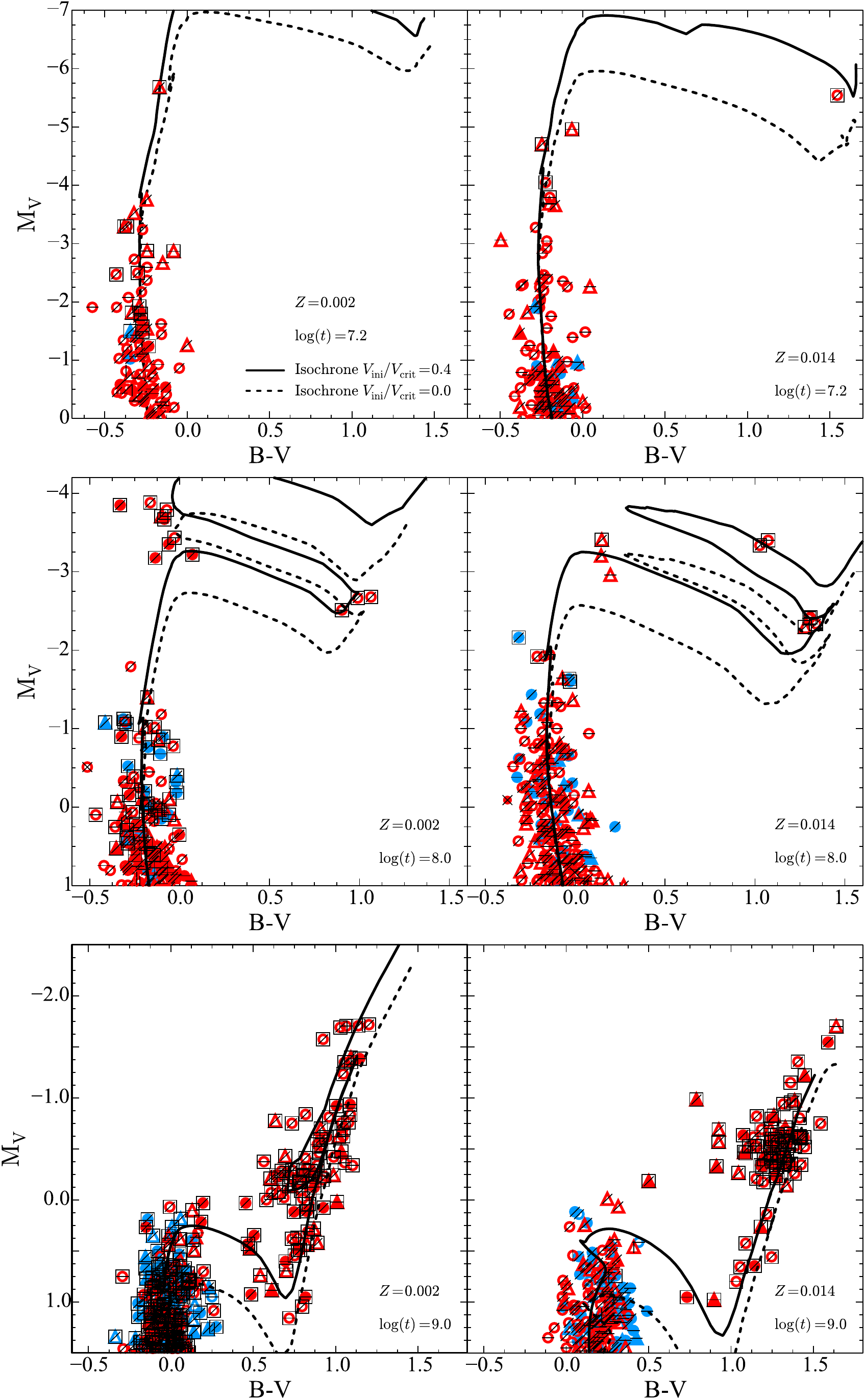}
\caption{Synthetic CMD for various ages and metallicities. The symbols are the same as in Fig.~\ref{CMD1}.}
\label{CMDZ}
\end{figure*}

\begin{table}[t]
\caption{Comparison of the counts of stars between the cluster presented in Fig.~\ref{CMD1} and the one in Fig.~\ref{CMDH}.}
\begin{center}
\scalebox{.9}{%
\begin{tabular}{c|ccc|ccc}
\hline\hline
\rule[0mm]{0mm}{5mm}M$_\text{V}$  & \multicolumn{3}{c|}{\citet{Huang2006a}} & \multicolumn{3}{c}{\citet{Huang2010a}} \\
\rule[-3mm]{0mm}{5mm}interval & $N^\star$ & $N^\star(0.85)$ &  & $N^\star$ & $N^\star(0.85)$ & \\
\hline
$[-2.25,-1]$	&	$21$	&	$7$	&	$33.3\%$	&	$19$	&	$6$	&	$31.6\%$ \rule[0mm]{0mm}{3mm}\\
$[-1,0]$	&	$40$	&	$3$	&	$7.5\%$	&	$69$	&	$13$	&	$18.8\%$ \\
\hline
\end{tabular}}
\end{center}
\label{tabla1}
\end{table}

Table~\ref{tabla1} presents the counts of MS stars in a given magnitude interval for the clusters computed with each velocity distribution. $N^\star(0.85)$ counts the number of stars rotating with $\omega>0.85$. In the M$_\text{V}$ interval $[-2.25,-1]$, the counts are rather similar for the two distributions, and thus the fraction of rapid rotators is about the same ($\sim30\%$). In the interval $[-1,0]$, the fraction of rapid rotators drops for both distribution. However, the drop is less steep for the \citet{Huang2010a} distribution \citep[$18.8\%$ against $7.5\%$ for][]{Huang2006a}. Indeed, stars with a mass below $4\ M_\sun$ appear in this magnitude interval, and in the \citet{Huang2010a} work, the velocity distribution for the mass range $2-4\ M_\sun$ peaks at higher velocity than the distribution for higher masses. 

\subsubsection{Synthetic colour-magnitude diagrams for different ages and metallicities}

Figure~\ref{CMDZ} presents the CMD of various clusters for different ages and metallicities. The symbols are the same than in Fig.~\ref{CMD1}.

The turnoff of the isochrone is brighter at $Z=0.014$ than at $Z=0.002$. Since stars at lower metallicity are more luminous, we would expect the turnoff to be brighter. However, they have also shorter lifetimes, so the mass at the turnoff at a given age is lower at $Z=0.002$ ($M_\text{TO,100\,Myr}=4.8\,M_\sun$) than at $Z_\sun$ ($M_\text{TO,100\,Myr}=5.2\,M_\sun$). For clusters, it is less clear, because velocity and angle dispersions as well as noise blur the picture.

Generally, the fraction of rapid rotators among stars located one magnitude below the turnoff increases with increasing age. At $\log(t)=7.2$, there are no rapid rotators brighter than M$_\text{V}=-2$. At $\log(t)=8.0$, the fraction is around $30-35\%$. At $\log(t)=9.0$ and $Z_\sun$, this trend continues ($45\%$). For the same age at $Z=0.002$, the minimal mass of stars one magnitude below the turnoff drops below $1.7\,M_\sun$, which is the minimal mass for the cluster, so the value obtained ($28\%$) is not reliable. There is no clear trend with metallicity for the two ages where the populations are complete.

At $Z_\sun$, the fraction of stars with $\text{(N/C)/(N/C)}_\text{ini}\geq3$ decreases with age. The opposite trend applies at $Z=0.002$. This is expected since at low $Z$, even low-mass stars get a surface enrichment strong enough to reach values higher than 3, while at solar metallicty, the lowest mass never reach such a level \citep[see Fig.~8 in][]{Georgy2013a}. As time proceeds, the low-mass stars become the dominant population in clusters, explaining why the fraction of stars with  $\text{(N/C)/(N/C)}_\text{ini}\geq3$ increases with age at $Z=0.002$.

As expected \citep[see][]{Georgy2013b}, the evolved stars are shifted to bluer colours in low metallicity clusters.

For young clusters, it is very difficult to determine an age from isochrones in M$_\text{V}$ vs B-V based only on the turnoff, because the stars remain at constant B-V for a while after the MS, extending from the turnoff\footnote{\footnotesize{Note that this problem does not arise with CMDs in M$_\text{B}$ vs U-B.}}. For older clusters, the crossing of the CMD occurs immediately after the turnoff, making the age determination more obvious. In all cases (but particularly for young clusters), the presence of evolved stars is a great help to constrain the best fitting isochrone. 

\subsection{On the evolution of stellar count}
 
\begin{figure}
  \centering
    \includegraphics[width=0.50\textwidth]{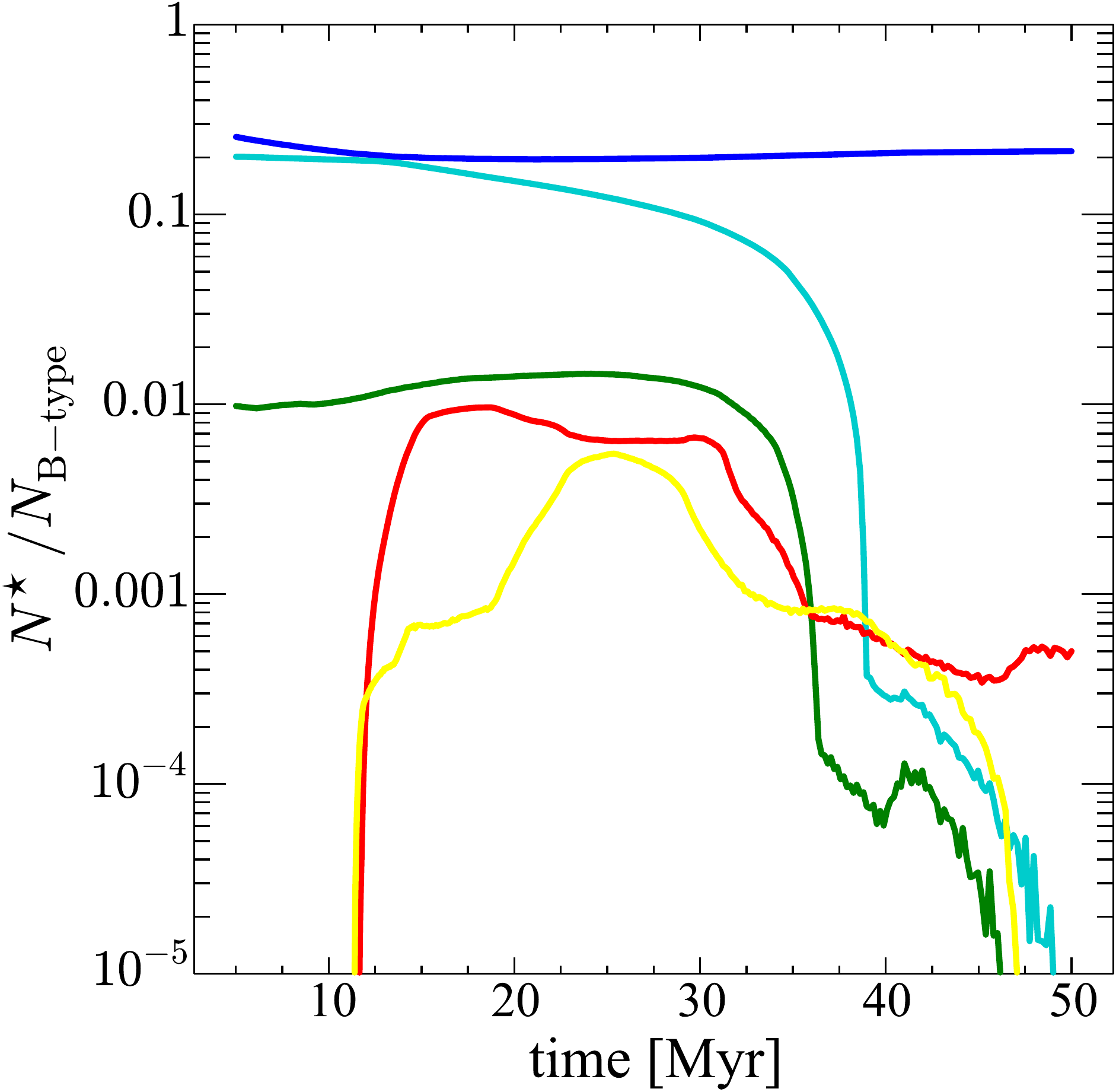}
\caption{Time evolution of different kinds of stellar populations, normalised to the total amount of B type stars at each time: B-type stars rotating with $\omega\geq0.80$ (blue line), early B-type stars (cyan line) and early B-type stars rotating with $\omega\geq0.80$ (green line), RSGs (red line), and YSGs (yellow line).}
\label{PopEvol}
\end{figure}

Figure~\ref{PopEvol} shows the time evolution of the fraction of B stars with a rotational rate $\omega$ higher than $0.8$, the fraction of early B stars with rate $\omega\geq0.80$, as well as the RSG and yellow supergiants (YSG)\footnote{Supergiant stars are defined here to be stars with $\log(L/L_\sun)>4.0$. Among them, RSG are defined to have $\log(T_\text{eff})<3.66$, and YSG $\log(T_\text{eff}$) between 3.66 and 3.9.}, normalised to the total number of B stars. This figure, corresponding to $Z=0.014$, shows that the observation of different stellar groups could be helpful to constrain the age of a cluster\footnote{\footnotesize{Because we do not account for stars more massive than 15 $M_\sun$, our results are valid for clusters older than $10\,\text{Myr}$ at solar metallicity, \textit{i.e.} when the turnoff point is in the range of the earliest B type stars.}}. It is unlikely to find RSG stars in clusters with an age below $14$ or above $30\,\text{Myr}$. They are expected to be found only in massive clusters, with hundreds of B-type stars. Such clusters are also expected to host tens of rapidly rotating B-type stars, as observed by \citet{Marco2013a} in the case of the massive star cluster NGC~7419. In less massive clusters, RSG stars are expected to be very rare. In general, RSGs will be always more numerous than YSGs.

This mode of ``star count evolution'' for SYCLIST was also used in \citet{Granada2013a} to study the time-evolution of Be populations in single-aged clusters.

\section{Application of SYCLIST to \object{NGC 663} - a Be star-rich open cluster hosting red giant stars}\label{Sec_NGC663}

Even though it is beyond the scope of the present article to make an extensive analysis of observed stellar populations, we present in this section an example demonstrating the potential of SYCLIST in the study of different stellar populations and dating of open clusters.

\citet{Pigulski2001a} presented BV(RI)$_c$H$_{\alpha}$ of the central region of the open cluster NGC 663, covering 14 $\times$ 20 arcmin$^2$. Their H$_{\alpha}$ photometry, complete down to magnitude R$_c$=15.4 (corresponding to A5 spectral type for cluster members), allowed them to detect all the stars presenting H$\alpha$ emission in B-type range, and to distinguish non-cluster stars which contaminate the field of NGC 663. They identified 392 cluster members with R$_c\leq15.4$, from which 26 are H$\alpha$ emitters of spectral type B, identified thus as Be stars. 

\begin{figure*}
\centering{
\includegraphics[width=0.45\textwidth]{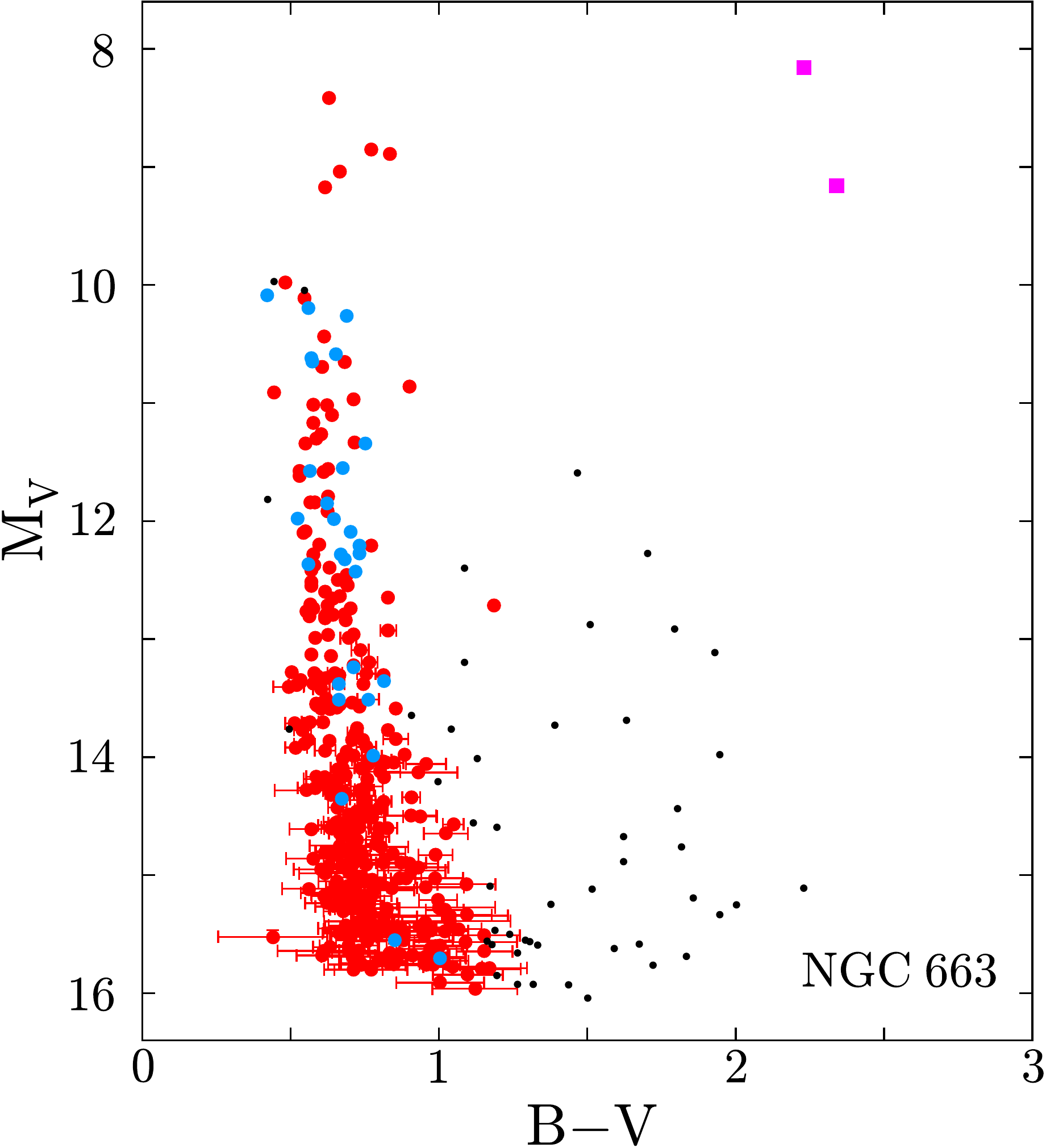}
\includegraphics[width=0.45\textwidth]{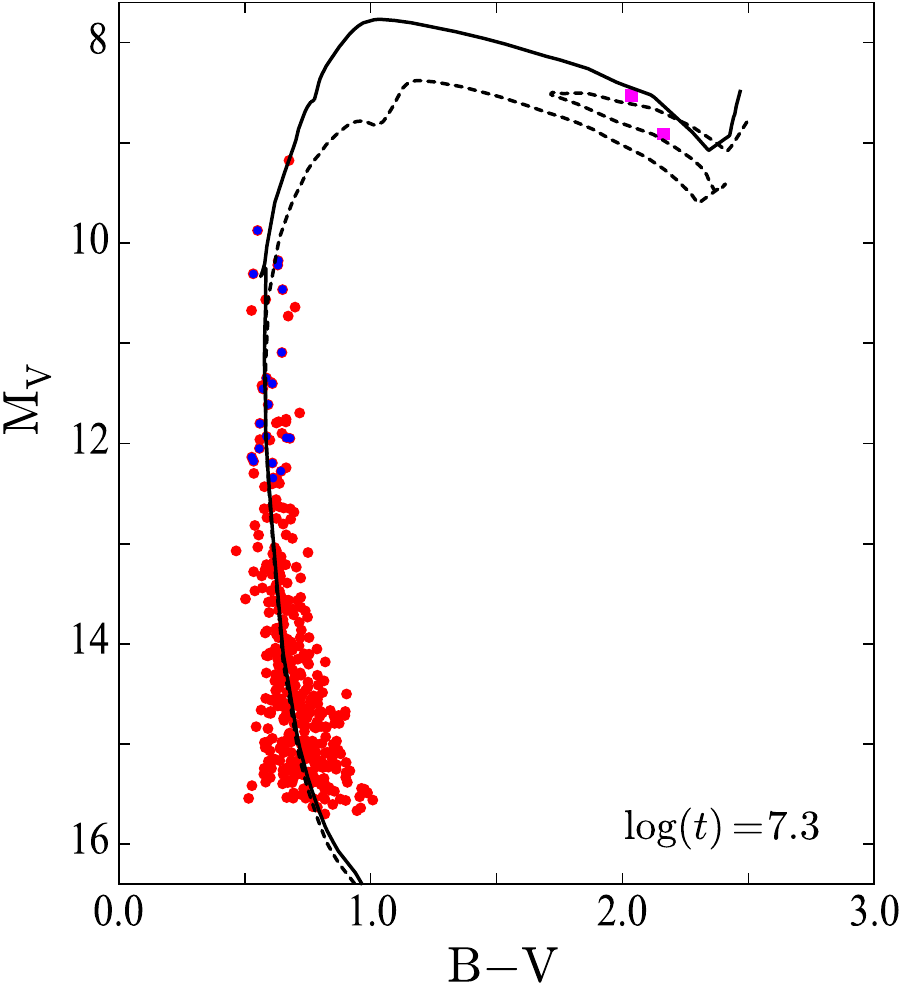}}
\caption{\textit{Left panel:} CMD of NGC 663. Black dots are the field stars from \citet{Pigulski2001a}. Red circles are cluster members with H$\alpha$ photometry, blue circles are Be stars and magenta squares red giant stars that are confirmed cluster members \citep{Mermilliod2008a}. \textit{Right panel:} Synthetic cluster of an age of $\log(t)=7.3$. Red points indicate stars with $T_\text{eff}$ higher than 10000\,K. Blue circles represent stars with $\omega>0.8$ and magenta squares stars with $\log(T_\text{eff})<3.66$. Isochrones at an age of $\log(t)=7.3$ are drawn ($V_\text{ini}/V_\text{crit}=0.40$: solid lines, non-rotating: dashed lines).}
\label{NGC_CMD}
\end{figure*}

Figure~\ref{NGC_CMD} (\textit{left panel}) shows the CMD of the stars in the above-mentioned field. We produced several synthetic clusters at various ages, assuming a \citet{Salpeter1955a} IMF and \citet{Huang2010a} initial velocity distribution. They contain the same amount of stars in the B-type range as observed in NGC 663 by \citet{Pigulski2001a}. We considered in this case that the errors in colour and magnitude increase with increasing V, as obtained from the observations by these authors. The cluster at $\log(t)=7.3$ ($20\,\text{Myr}$) appears to better describe the characteristics of NGC 663. This cluster is plotted in Fig.~\ref{NGC_CMD} (\textit{right panel}), assuming a normal extinction law, E(B-V)=0.83, and a true distance modulus of 11.6 mag \citep{Pigulski2001a}. Isochrones at $\log(t)=7.3$, with and without rotation, are also shown.

Different authors obtain different ages between 10 and 30 Myr for NGC 663 \citep[\textit{e.g.}][]{Pigulski2001a,Pandey2005a}. As mentioned in the previous section, the expected fraction of RSG reaches a maximum value around 18 Myrs. At this age, the fraction $N_\text{RSG}/N_\text{B-type}=1\%$ (see Fig.~\ref{PopEvol}). This implies an expected number of $2-6$ RSGs in such a cluster. The observed value ($2$) is well in that range. NGC 663 exhibits also a relatively large fraction of Be stars below the turnoff. However, the Be stars at higher magnitudes are not expected in the framework of our models. At detailed study is deferred to a forthcoming paper.

NGC 663 exhibits also a relatively large fraction of rapid rotators, which is well in line with the findings of the previous section.

Even though our synthetic cluster has a remarkable resemblance to the observed one in Figure \ref{NGC_CMD}, and the study of the evolution of stellar populations allows us to understand the presence of different stellar populations in the CMD at different ages, such as RSG stars or populations of rapidly rotating stars, there are some observed features that remain uncertain. The presence of blue supergiant stars is not an unusual feature in a cluster of this age, and 5 such BSGs are observed in NGC 663. However, their presence cannot be explained with our single stellar population synthesis. They could originate from the merger of stars in binary systems, or be rapidly rotating more massive stars that are not accounted for in this study \citep[see][]{Schneider2014a}.

\section{Conclusions}\label{Sec_Conclu}

We present the SYCLIST code, a new tool for interpolating between stellar tracks, building isochrones, creating synthetic clusters, and following the evolution of stellar populations. It includes an IMF, various initial velocity and viewing angle distributions, and is able to account for the gravity- and limb-darkening. The binary fraction and a photometric noise are additional options for the outputs of the ``Synthetic cluster'' mode.

In this paper we explain how these effects are implemented in the code, and study to which extent they impact the aspects of stellar tracks, isochrones, and synthetic clusters. We also study typical synthetic clusters at various ages and metallicities, and discuss their main features.

We briefly present the potential use of the SYCLIST code in comparison with observed clusters. More extensive studies and comparisons will be the subject of a forthcoming paper.

\begin{acknowledgements}
The authors thank the anonymous referee for her/his positive comments and constructive suggestions. CG acknowledges support from the European Research Council under the European Union's Seventh Framework Programme (FP/2007-2013) / ERC Grant Agreement n. 306901.
\end{acknowledgements}

\bibliographystyle{aa}
\bibliography{MyBiblio}
\listofobjects
\end{document}